\documentclass[twocolumn, aps, prc, superscriptaddress, floatfix]{revtex4}

\usepackage{multirow}
\usepackage{graphics}
\usepackage{epsfig}
\usepackage{amsmath}

\usepackage{subfigure}
\usepackage{latexsym}
\usepackage{feynmp}
\usepackage{float}

\usepackage[normalem]{ulem}  
\usepackage[dvips]{color} 

\renewcommand\sout{\bgroup \color{red} \ULdepth=-.5ex \ULset}

\begin{document}

\title{Effects of charm-strange resonances on charmed meson production by recombination in relativistic heavy ion collisions}

\author{Sungtae Cho}\email{sungtae.cho@kangwon.ac.kr}
\affiliation{Division of Science Education, Kangwon National
University, Chuncheon 24341, Korea}
\author{Su Houng Lee}\email{suhoung@yonsei.ac.kr}
\affiliation{Department of Physics and Institute of Physics and
Applied Physics, Yonsei University, Seoul 03722, Korea}

\begin{abstract}

We discuss charmed mesons in heavy-ion collisions by focusing on
their production from the quark-gluon plasma by quark
recombination. We consider the production of not only the $D$,
$D^*$, $D_s$, and $D_s^*$, but also the $D_{s0}^*(2317)$,
$D_{s1}(2460)$, $D_{s1}(2536)$, and $D_{s2}^*(2573)$, thereby
investigating the important role of heavier charm-strange
resonances in understanding the measurements of the $D$ and $D_s$
mesons at the Large Hadron Collider at $\sqrt{s_{NN}}=5.02$ TeV.
We calculate their yields and transverse momentum distributions,
and also evaluate ratios of the transverse momentum distribution
between various charmed mesons, i.e., the $D_s$ and $D^0$, the
$D_s$ and $D^+$, $D_s^*$ and $D^{*0}$, and $D_s^*$ and $D^{*+}$.
Through this analysis, we attempt to understand in more detail the
effects of strangeness enhancement on charmed meson production in
heavy-ion collisions at $\sqrt{s_{NN}}=5.02$ TeV.

\end{abstract}



\maketitle

\section{Introduction}

Relativistic heavy-ion collision experiments have served as
indispensable laboratories for studying a system of quantum
chromodynamic matter under extreme conditions, the quark-gluon
plasma (QGP) \cite{Adams:2005dq, Adcox:2004mh, Gyulassy:2004zy}.
As useful probes for investigating the properties of the QGP,
charmed hadrons have attracted increasing attention as the
collision energies available in heavy-ion experiments have
increased. The suppression of the $J/\psi$ production, owing to
color screening between quarks caused by s system composed of
deconfined quarks and gluons, has been suggested as one of the
signatures for the existence of the QGP in heavy-ion collision
experiments \cite{Matsui:1986dk}. Due to the enormous energies
currently available in relativistic heavy-ion collisions at the
Large Hadron Collider (LHC), many more charmed hadrons are
observed nowadays, thereby providing valuable opportunities to
study the QGP in more detail \cite{Thews:2000rj, Andronic:2007bi}.

The enhanced transverse momentum distribution ratio between the
baryon and meson in the intermediate transverse momentum region
for charmed hadrons, $\Lambda_c/D^0$, has been studied
\cite{Oh:2009zj, Plumari:2017ntm, Cho:2019lxb}, providing evidence
for the role of quark coalescence in charmed-hadron production
from the QGP. The production of charmed exotic hadrons
\cite{Cho:2010db, Cho:2011ew, Cho:2017dcy} or multicharmed hadrons
\cite{Cho:2019syk, Minissale:2023dct, Andronic:2021erx} in heavy
ion collisions has also been investigated, presenting the
possibilities for observing those hadrons for the first time in
nature and thus providing insight into the production mechanism
involving charm quarks.

Nevertheless, the overall understanding of the charmed hadron
production in heavy-ion collisions is rather limited due to
insufficient knowledge of the properties of the charmed hadrons
themselves. The charm quark is known to be produced predominantly
in the initial hard scatterings in heavy-ion collision
experiments, since the energy available in the later stages of the
collision is generally insufficient for its production due to its
large mass. After it is produced, its number is conserved
throughout the entire process in heavy ion collisions, even after
the hadronization stage. Therefore, it is essential to consider
the production of all charmed hadrons in order to understand the
production of any specific charmed hadron, as they are all related
in decay processes as mother and daughter hadrons. More
importantly, they are also connected by sharing the finite number
of charm quarks as the source of their constituents in their
production.

Thus, the current explanation of the production of charmed hadrons
is considered to be the best given the limited information
available on charmed hadrons, and there is still room for
improvement in explaining the charmed hadron production, e.g., the
enhancement in the transverse momentum distribution ratio between
the $\Lambda_c$ and $D^0$.

Regarding the same ratio in the light hadrons, $\bar{p}/\pi$, both
the direct production of the proton and pion from the QGP and
their indirect production from well-established resonances are
already large.  As a result, the ratio is not expected to be
strongly affected by the inclusion of the much heavier resonances
or by the observation of new heavier resonances that decay into
the proton and pion. Moreover, the mass difference between the
proton and the lightest resonances, the delta, is already more
than a 30$\%$ of the proton mass, making the feed-down
contributions from heavier resonances negligible.

On the other hand, the yields of heavier resonances decaying to
charmed mesons are not negligible compared to those of the $D$ and
$D_s$, and thus the consideration of heavier resonance production
leads to large deviations in the production ratio between the
$\Lambda_c$ and $D^0$, compared to the experimental measurements
in heavy ion collisions, highlighting the importance of carefully
examining their contributions. It has been shown that the
production of the heavier resonances for the $D_s$ meson, e.g.,
the $D_{s0}^*(2317)$ and $D_{s1}(2460)$, is comparable to the
production of the $D_s$, responsible for about 30$\%$ of the total
$D_s$ yield \cite{Cho:2025lrc}.

The production of non-strange charmed mesons, e.g., the $D$ and
$D^*$, has also not been fully understood so far. For example, the
estimated yields of the $D^+$ and $D^{*+}$ in the Statistical
Hadronization Model with charm (SHMc) \cite{Andronic:2021erx}
deviates significantly from the corresponding experimental
measurements by the ALICE Collaboration at the LHC
\cite{ALICE:2021rxa}. This is likely attributable to the limited
information on the production of heavier charmed mesons; there are
many scarcely known charmed strange or non-strange charmed mesons
with masses below 3 GeV \cite{ParticleDataGroup:2024cfk} that are
expected to decay mostly to light charmed mesons, $D$ and $D^*$
mesons.


For this reason, we investigate here the production of not only
the $D$, $D^*$, $D_s$, and $D_s^*$, but also four charm-strange
mesons, i.e., the $D_{s0}^*(2317)$, $D_{s1}(2460)$,
$D_{s1}(2536)$, and $D_{s2}^*(2573)$, focusing on their production
from the quark-gluon plasma by quark recombination based on the
assumption of the sudden hadronization between constituents in the
QGP \cite{Greco:2003xt, Greco:2003mm, Fries:2003vb, Fries:2003kq}.
Currently, explicit information on the decay channels is available
only for the above eight mesons as far as charmed mesons are
concerned. Nevertheless, the decay channels are still known
partially for the $D_{s1}(2460)$ and $D_{s2}^*(2573)$: only 82$\%$
and 37$\%$ of the branching ratios are available among all the
decay modes of the $D_{s1}(2460)$ and $D_{s2}^*(2573)$,
respectively \cite{ParticleDataGroup:2024cfk}.

The inclusion of the above four charm-strange mesons in the
discussion of charmed hadron production affects the production of
not only charm-strange mesons but also non-strange charmed mesons.
This is because two of the charm-strange mesons, the
$D_{s0}^*(2317)$ and $D_{s1}(2460)$ decay exclusively to light
charm-strange mesons, whereas the other two charm-strange mesons,
the $D_{s1}(2536)$ and $D_{s2}^*(2573)$ decay mainly to
non-strange charmed mesons \cite{ParticleDataGroup:2024cfk}. Thus,
by evaluating the transverse momentum distributions and yields of
the above charmed mesons, we can gain a better understanding of
the production of the $D^0$, $D^+$, $D^{*+}$ and $D_s$ measured at
LHC at $\sqrt{s_{NN}}=5.02$ TeV.

The production of charmed mesons is important as it also affects
the production of other hadrons, e.g., the exotic hadrons in
hadronic molecular states made up of charmed mesons such as the
presumed hadronic molecular state of the $X(3872)$ composed of the
$D^0$ and $\bar{D}^{*0}$ or the $\bar{D}^0$ and $D^{*0}$.
Depending on the yield of charmed mesons, the expected yield of
the $X(3872)$ meson in a hadronic molecular state would also vary,
leading to different conclusions about its internal structure
\cite{Cho:2010db, Cho:2011ew, Cho:2017dcy}.

In our analysis, we take into account the conservation of the
total number of charm quarks in the entire process in heavy ion
collisions after their initial production in the hadronic
scattering. This is accomplished not only by enforcing charm
quarks in low transverse momentum regions to be hadronized into
charmed hadrons by quark recombination, but also by converting all
the remaining charm quarks to charmed hadrons by fragmentation,
mostly in high transverse momentum regions. Thus, the inclusion of
new heavier charmed resonances leads to the reduction in the
yields of all other charmed hadrons compared to the previous
analysis \cite{Cho:2025lrc}, while the yields of hadrons which can
be produced indirectly from these new heavier charmed resonances
are expected to increase.

We also evaluate the transverse momentum distribution ratios
between various charmed mesons, i.e., the $D_s$ and $D^0$, the
$D_s$ and $D^+$, $D_s^*$ and $D^{*0}$, and $D_s^*$ and $D^{*+}$,
attempting to understand in more detail the effects of the
strangeness enhancement on the charmed meson production in
heavy-ion collisions at $\sqrt{s_{NN}}=5.02$ TeV. Moreover, we
compare the production of heavier resonances to that of the
charmed mesons, i.e., the $D_{s0}^*(2317)$ and $D_{s1}(2460)$ to
the $D_s^{*}$ and the $D_{s1}(2536)$ and $D_{s2}^*(2573)$ to the
$D^{*+}$, thereby presenting an efficient way of measuring these
heavier resonances in heavy ion collisions.

The paper is organized as follows. In Sec. II, we consider the
production of charmed mesons in relativistic heavy ion collisions,
introducing the transverse momentum distribution of charmed mesons
in the coalescence model. We discuss in Sec. II transverse
momentum distributions of constituent quarks, and present the
transverse momentum distribution of charm quarks obtained by
referring to the measurement of that of the $D^0$ meson by ALICE
Collaboration at the LHC. Then, we calculate in Sec. III the
transverse momentum distributions and yields of not only the $D$,
$D^*$, $D_s$, and $D_s^*$, but also the $D_{s0}^*(2317)$,
$D_{s1}(2460)$, $D_{s1}(2536)$, and $D_{s2}^*(2573)$ meson, and
compare them to the measurements by ALICE Collaboration at the
LHC. Furthermore, we evaluate the transverse momentum distribution
ratios between various charmed hadrons in Sec. IV. Sec. V is
devoted to conclusions.

\section{Charmed meson production in heavy ion collisions}

We consider here charmed meson production from charm and light
quarks, e.g., up, down and strange quarks in the quark-gluon
plasma by recombination. Based on the coalescence model
\cite{Fries:2003kq, Fries:2003vb, Greco:2003xt, Greco:2003mm} we
evaluate the yields and transverse momentum distributions of $D$,
$D^*$, $D_s$, $D_s^*$, $D_{s0}^*(2317)$, $D_{s1}(2460)$,
$D_{s1}(2536)$, and $D_{s2}^*(2573)$ mesons.

The transverse momentum distribution of charmed mesons in the
non-relativistic limit is given by \cite{Greco:2003mm,
Greco:2003xt, Oh:2009zj, Cho:2025lrc},

\begin{eqnarray}
&& \frac{dN_M}{d\vec p_T}=\frac{g_M}{V} \int d\vec r d\vec
p_{cT}d\vec p_{\bar{q}T} \delta^{(2)}(\vec p_T-\vec p_{cT}-
\vec p_{\bar{q}T}) \nonumber \\
&& \qquad\quad\times\frac{dN_{c}}{d\vec p_{cT}}
\frac{dN_{\bar{q}}}{d \vec p_{\bar{q}T}} W_M(\vec r, \vec k),
\label{TransCharmed}
\end{eqnarray}
when charmed mesons are produced from one charm and one anti-light
quarks by recombination with the subscript $c$ and $\bar{q}$
representing the charm and anti-light quarks, respectively. In Eq.
(\ref{TransCharmed}), $W_M(\vec r,\vec k)$ and $g_M$ are the
Wigner function and the degeneracy factor of a charmed meson,
respectively. The subscript $M$ in Eq. (\ref{TransCharmed}) stands
for a charmed meson of interest: the $D$, $D^*$, $D_s$, $D_s^*$,
$D_{s0}^*(2317)$, $D_{s1}(2460)$, $D_{s1}(2536)$, and
$D_{s2}^*(2573)$.

In Eq. (\ref{TransCharmed}), the following relative coordinates
and transverse momenta have been adopted for the Wigner function,
\begin{eqnarray}
&& \vec R=\frac{m_{\bar{q}}\vec r_{\bar{q}}+m_c\vec r_c}
{m_{\bar{q}}+m_c}, \qquad \vec r=\vec r_{\bar{q}}-\vec r_c, \nonumber \\
&& \vec K=\vec p_{\bar{q}T}'+\vec p_{cT}', \qquad \vec
k=\frac{m_c\vec p_{\bar{q}T}'-m_{\bar{q}}\vec
p_{cT}'}{m_{\bar{q}}+m_c}, \label{rel_coordinates}
\end{eqnarray}
where $\vec p_{cT}'$ and $\vec p_{\bar{q}T}'$ are the transverse
momenta of charm and light quarks in the rest frame of charmed
mesons, Lorentz transformed from the transverse momenta of those
quarks in the fireball frame, $\vec p_{cT}$ and $\vec
p_{\bar{q}T}$, respectively \cite{Scheibl:1998tk, Oh:2009zj}.

We consider here two Wigner functions, one for charmed mesons in
$s$-wave states such as $D$, $D^*$, $D_s$ and $D_s^*$ mesons, and
the other for those in $p$-wave states, i.e., $D_{s0}^*(2317)$,
$D_{s1}(2460)$, $D_{s1}(2536)$, and $D_{s2}^*(2573)$ mesons,
\begin{eqnarray}
&& W_s(\vec r, \vec k) = 8 e^{-\frac{r^2}{\sigma^2}-k^2
\sigma^2} \nonumber \\
&& W_p(\vec r, \vec k) = \bigg( \frac{16}{3} \frac{r^2}{\sigma^2}
-8+\frac{16}{3} \sigma^2 k^2 \bigg) e^{-\frac{r^2}{\sigma^2}-k^2
\sigma^2}, \label{Wignersp}
\end{eqnarray}
built from wave function of charmed mesons using harmonic
oscillator wave functions. In Eq. (\ref{Wignersp})
$\sigma^2=1/(\mu\omega)$ with $\omega$ and
$\mu=m_{\bar{q}}m_c/(m_{\bar{q}}+m_c)$ being the oscillator
frequency of the harmonic oscillator wave function and the reduced
mass, respectively.

After carrying out the integration of the Wigner function, Eq.
(\ref{Wignersp}), over a relative coordinate, $\vec r$ in Eq.
(\ref{TransCharmed}) one obtains \cite{Cho:2025lrc},

\begin{eqnarray}
&& \frac{dN_M}{d\vec p_T}=\frac{g_M}{V}(2\sqrt{\pi}\sigma)^3
\int d\vec p_{cT}d\vec p_{\bar{q}T}e^{-\sigma^2 k^2} \nonumber \\
&& \qquad\quad \times\delta^{(2)}(\vec p_T-\vec p_{cT}-\vec
p_{\bar{q}T}) \frac{dN_c}{d\vec p_{cT}}\frac{dN_{\bar{q}}}{d \vec
p_{\bar{q}T}}, \label{CoalTransCharmed}
\end{eqnarray}
for the $D$, $D^*$, $D_s$ and $D_s^*$, and
\begin{eqnarray}
&& \frac{dN_M}{d\vec p_T}=\frac{g_M}{V}(2\sqrt{\pi}\sigma)^3 \int
d\vec p_{cT}d\vec p_{\bar{s}T}\Big(\frac{2}{3}
\sigma^2k^2\Big) e^{-\sigma^2 k^2} \nonumber \\
&& \qquad\quad \times\delta^{(2)}(\vec p_T-\vec p_{cT}-\vec
p_{\bar{s}T}) \frac{dN_c}{d\vec p_{cT}}\frac{dN_{\bar{s}}}{d \vec
p_{\bar{s}T}}, \label{CoalTransCSp}
\end{eqnarray}
for the $D_{s0}^*(2317)$, $D_{s1}(2460)$, $D_{s1}(2536)$, and
$D_{s2}^*(2573)$.

We adopt here the following transverse momentum distribution for
light quarks at mid-rapidities in central collisions,

\begin{eqnarray} && \frac{dN_q}{d \vec p_{qT}}=\left\{
\begin{array}{l}
\frac{V}{(2\pi)^3}m_Te^{-m_T/T_{eff}}, \quad~~~p_{qT} \le 1.50~\textrm{GeV} \\
21.95 ~\mathrm{(GeV^{-2})} e^{-0.17(p_{qT}/p_{0T})^{3.23}} \\
+\frac{80112 ~\mathrm{(GeV^{-2})}}
{(1.0+(p_{qT}/p_{0T})^{0.65})^{10.29}}, ~p_{qT} >1.50~\textrm{GeV}
\end{array} \right. \nonumber \\
\label{d2NqdpT2}
\end{eqnarray}
with $g_q$, $V$, $m_T=\sqrt{p_{T}^2+m_q^2}$ and $T_{eff}$ being
the degeneracy factor for the spin and color of light quarks, the
coalescence volume, the transverse mass and the effective
temperature, respectively. The $p_{0T}$ in Eq. (\ref{d2NqdpT2}) is
taken to be 1.0 GeV. As the baryon chemical potential at the LHC
$\sqrt{s_{NN}}=5.02$ TeV energy is almost negligible, the
transverse momentum distribution of anti-quarks is assumed to be
same as that of quarks. The transverse momentum distribution, Eq.
(\ref{d2NqdpT2}) with the mass and effective temperature of
strange quarks is initially obtained such that it explains the
measured transverse momentum distribution of the $\phi$ meson at
$\sqrt{s_{NN}}=5.02$ TeV in heavy ion collisions
\cite{Cho:2025lrc}.

In regard to the transverse momentum distribution of charm quarks,
we modify that in Ref. \cite{Cho:2025lrc} after taking into
account the production of two more charm-strange mesons,
$D_{s1}(2536)$ and $D_{s2}^*(2573)$. Thus, we consider eight charm
mesons, or $D$, $D^*$, $D_s$, $D_s^*$, $D_{s0}^*(2317)$,
$D_{s1}(2460)$, $D_{s1}(2536)$, and $D_{s2}^*(2573)$ and ten charm
baryons, $\Lambda_c$, $\Sigma_c(2455)$, $\Sigma_c(2520)$,
$\Lambda_c(2595)$, $\Lambda_c(2625)$, $\Xi_c$, $\Xi_c'$,
$\Xi_c(2645)$, $\Omega(2695)$, and $\Omega(2770)$, and we require
the sum of the yields of above 18 charm hadrons as functions of
charm quark transverse momenta is equal to that of charm quarks
produced in initial hard collisions at zero transverse momentum,
meaning that all charm quarks at zero transverse momentum is
converted entirely into charmed hadrons by recombination
\cite{Oh:2009zj, Cho:2019syk, Cho:2025lrc}.

At the same time, we find the transverse momentum distribution of
charm quarks to explain that of $D^0$ mesons produced from charm
and light quarks in the quark-gluon plasma by coalescence,
measured in mid-rapidities at central collisions in relativistic
heavy ion collisions at $\sqrt{s_{NN}}=5.02$ TeV by ALICE
Collaboration \cite{ALICE:2021rxa}. In evaluating the transverse
momentum distribution of $D^0$ mesons, we have also taken into
account the contribution from the production of $D^0$ mesons by
fragmentation in high transverse momentum regions. All remaining
charm quarks after the consumption of charm quarks by coalescence
have been converted into charmed hadrons based on the Peterson
fragmentation function \cite{Peterson:1982ak} with
$\epsilon_c=0.022$. About 60.86$\%$ of the remaining charm quarks
are used to produce $D^0$ mesons via fragmentation when all
feed-down contributions are included \cite{Lisovyi:2015uqa}.

We show below the transverse momentum distribution of charm quarks
in mid-rapidities at 0-10$\%$ centrality in Pb-Pb collisions at
$\sqrt{s_{NN}}=5.02$ TeV,


\begin{widetext}

\begin{figure}[!h]
\begin{center}
\includegraphics[width=0.49\textwidth]{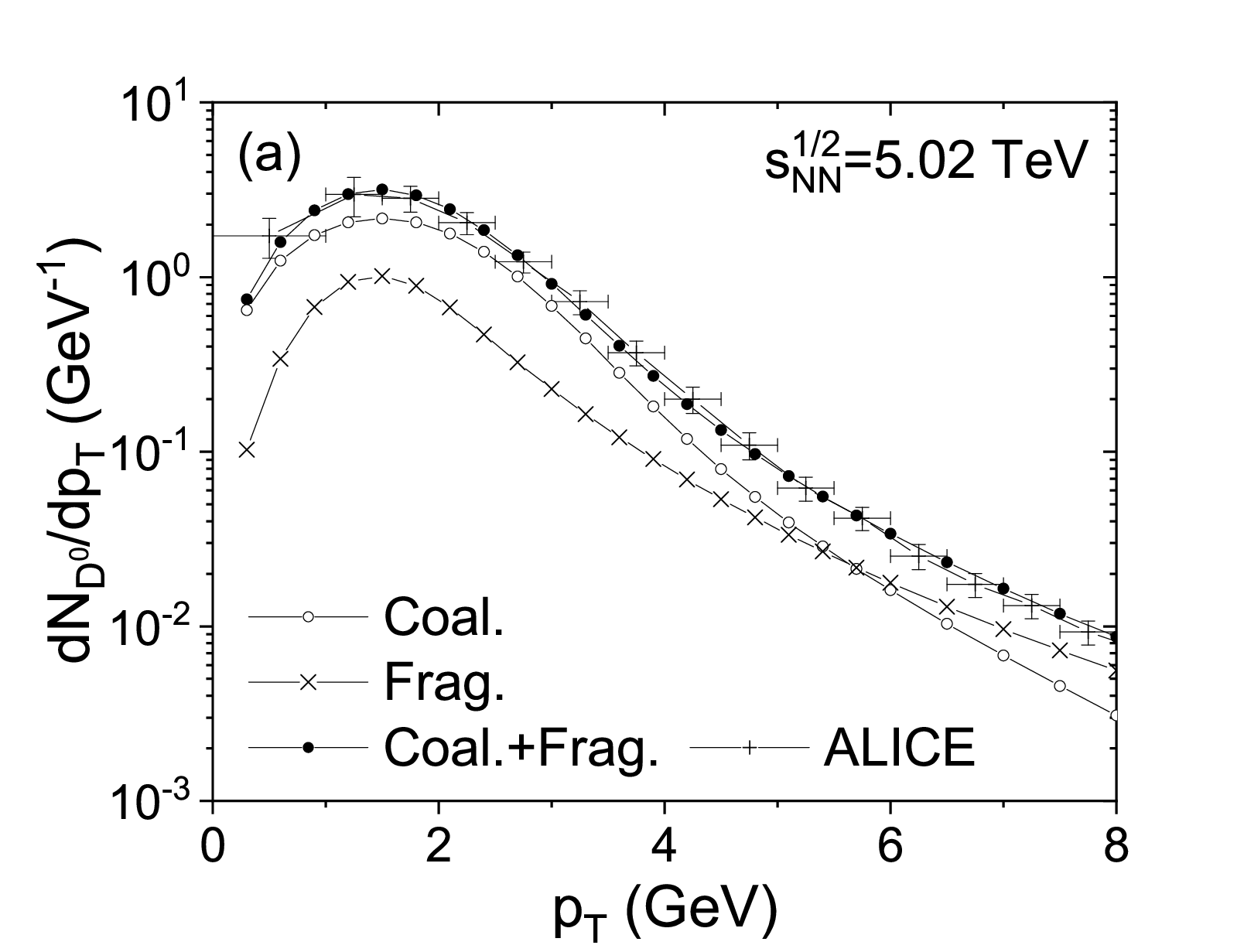}
\includegraphics[width=0.49\textwidth]{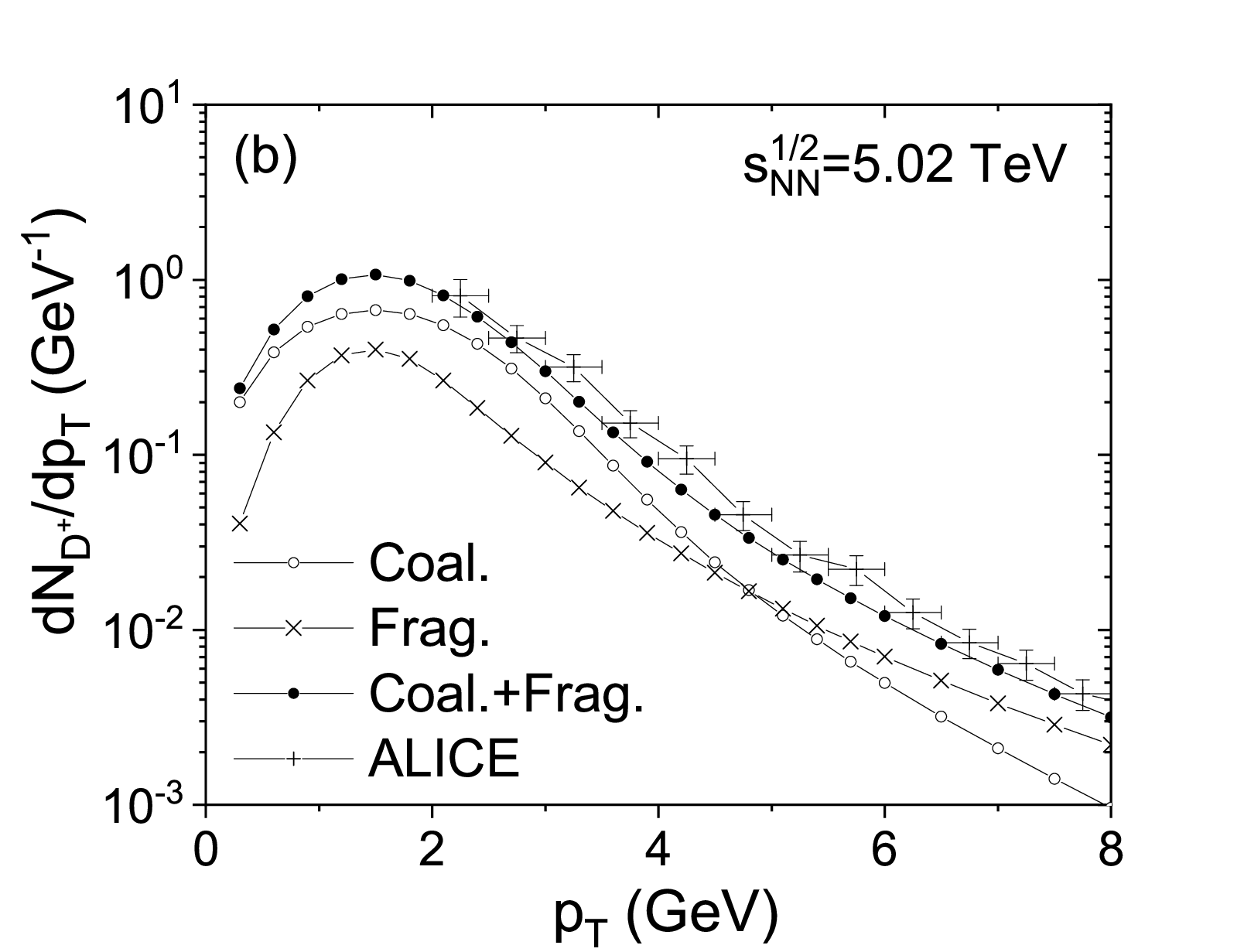}
\includegraphics[width=0.49\textwidth]{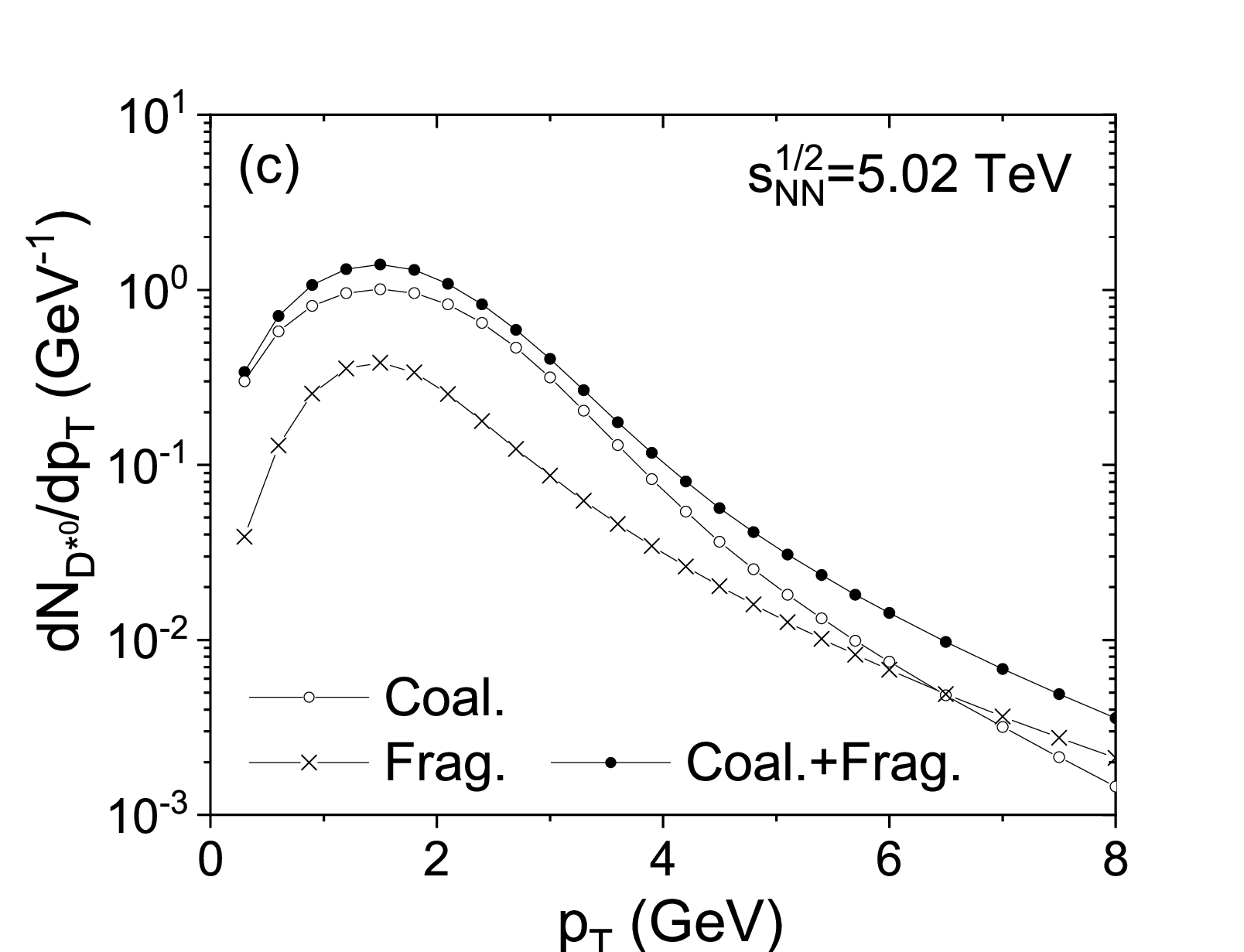}
\includegraphics[width=0.49\textwidth]{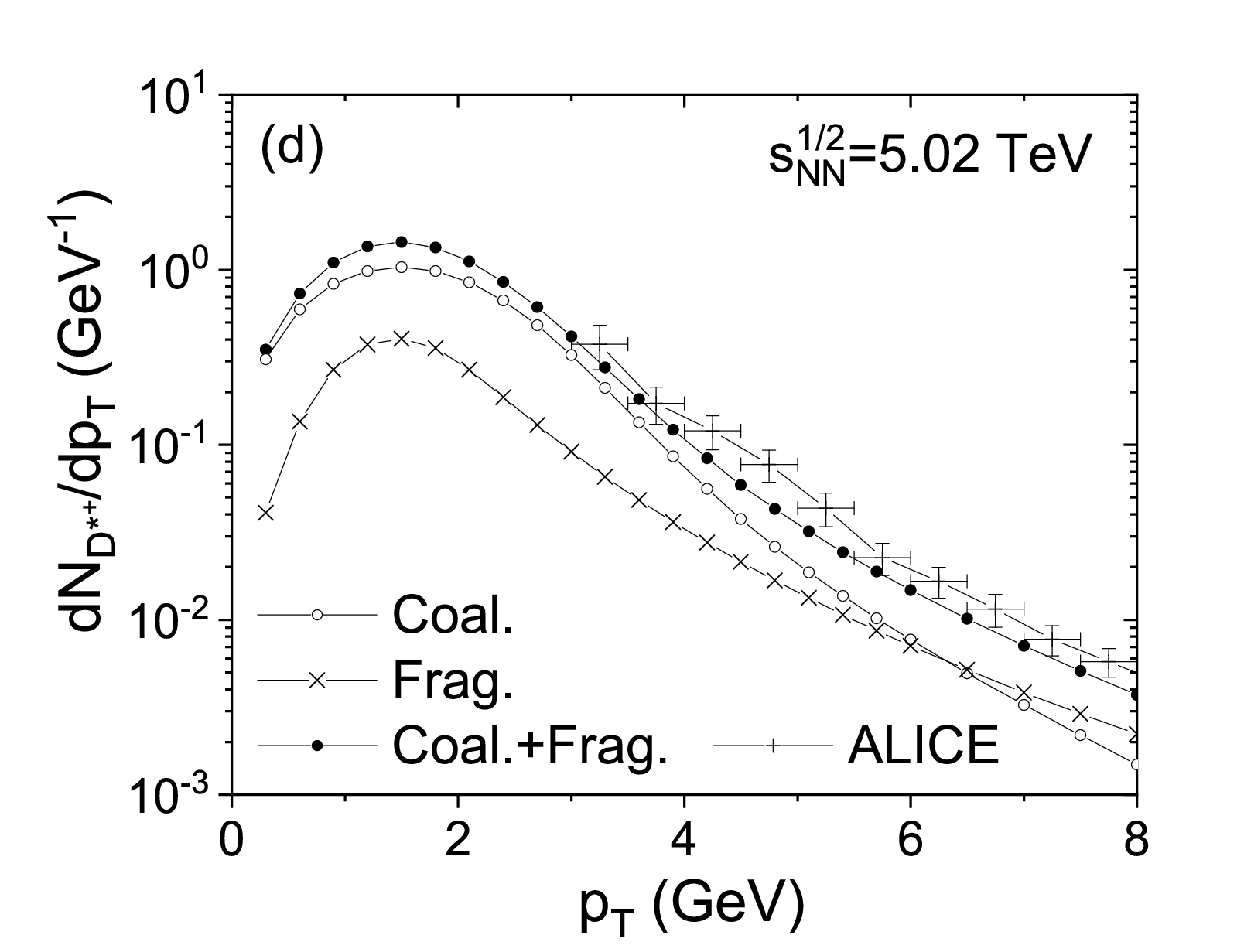}
\end{center}
\caption{Transverse momentum distribution of the $D^0$ (a), $D^+$
(b), $D^{*0}$ (c) and $D^{*+}$ (d) at mid-rapidity,
$dN_{D^0}/dp_T$, $dN_{D^+}/dp_T$, $dN_{D^{*0}}/dp_T$ and
$dN_{D^{*+}}/dp_T$ at the LHC $\sqrt{s_{NN}}=5.02$ TeV energy,
respectively, together with the experimental measurement of those
for the $D^0$, $D^+$ and $D^{*+}$ by ALICE Collaboration
\cite{ALICE:2021rxa} at 0-10$\%$ centrality in Pb-Pb collisions at
$\sqrt{s_{NN}}=5.02$ TeV. Also shown are transverse momentum
distributions of $D^0$, $D^+$, $D^{*0}$ and $D^{*+}$ mesons
produced purely by coalescence and also by fragmentation,
respectively. Here, feed-down contributions to the production of
$D^0$, $D^+$, $D^{*0}$ and $D^{*+}$ mesons, Eq.
(\ref{DDvfeeddown}) have been taken into account. }
\label{DDvpTdistribution}
\end{figure}

\end{widetext}

\begin{eqnarray}
&& \frac{d^2N_c}{d^2\vec p_{cT}}=\left\{
\begin{array}{l}
1.63 ~\mathrm{(GeV^{-2})} e^{-0.29(p_{cT}/p_{0T})^{2.03}}, \\
\qquad\qquad\qquad\qquad\quad\quad~~ p_{cT} \le 1.80~\textrm{GeV} \\
7.95 ~\mathrm{(GeV^{-2})} e^{-3.49(p_{cT}/p_{0T})^{3.59}} \\
+\frac{90112
~\mathrm{(GeV^{-2})}}{(1.0+(p_{cT}/p_{0T})^{0.50})^{14.19}},
~p_{cT} > 1.80~\textrm{GeV}
\end{array} \right.  \label{d2NcdpT2}
\end{eqnarray}
resulting in its yield at mid-rapidity, $dN_c/dy=16.7$.

We take the effective temperature in Eq. (\ref{d2NqdpT2}),
$T_{eff}$=189 and 173 MeV with constituent quark masses, $m_l$=300
and $m_s$=500 MeV for light and strange quarks, respectively, and
adopt 3360 fm$^3$ for the coalescence volume at
$\sqrt{s_{NN}}=5.02$ TeV at the LHC \cite{Cho:2025lrc}. The
oscillator frequency in the Wigner function,
$\omega=1/\mu\sigma^2$, has been determined, as mentioned before,
on the required condition that the yield of charmed hadrons is
equal to the number of charm quarks produced by initial hard
collisions at zero transverse momentum, yielding $\omega=0.119$
GeV.

\section{Transverse momentum distributions and yields of
charmed mesons}

With the transverse momentum distribution of charm and anti-light
quarks, Eqs. (\ref{d2NqdpT2}) and (\ref{d2NcdpT2}) we compute the
transverse momentum distributions of charmed mesons, or the $D$,
$D^*$, $D_s$, $D_s^*$, $D_{s0}^*(2317)$, $D_{s1}(2460)$,
$D_{s1}(2536)$, and $D_{s2}^*(2573)$ meson formed from charm and
anti-light quarks by coalescence in relativistic heavy ion
collisions at $\sqrt{s_{NN}}=5.02$ TeV at LHC.

We first show in Fig. \ref{DDvpTdistribution}(a) the transverse
momentum distributions of the $D^0$ produced by both recombination
and fragmentation, together with the sum of those two, or the
total transverse momentum distribution of the $D^0$ meson. The
experimental measurement of the $D^0$ transverse momentum
distribution by ALICE Collaboration at 0-10$\%$ centrality in
Pb-Pb collisions at $\sqrt{s_{NN}}=5.02$ \cite{ALICE:2021rxa} is
also shown in Fig. \ref{DDvpTdistribution}(a).

As shown in Fig. \ref{DDvpTdistribution}(a), the measurement of
the transverse momentum distribution for the $D^0$ meson by ALICE
Collaboration at 0-10$\%$ centrality in Pb-Pb collisions at
$\sqrt{s_{NN}}=5.02$ is well described by the transverse momentum
distribution of charm and anti-light quarks, Eqs. (\ref{d2NqdpT2})
and (\ref{d2NcdpT2}). The production of the $D^0$ by coalescence
is found to prevail in low transverse momentum regions, whereas
the $D^0$ is more dominantly produced by fragmentation in high
transverse momentum regions, as shown in Fig.
\ref{DDvpTdistribution}(a). In between the two production
mechanisms compete, and both contributions to production of the
$D^0$ are equal at about $p_T\approx 5.5$ GeV.

As mentioned before, it has been observed that the $D_{s1}(2536)$
and $D_{s2}^*(2573)$ decay to non-strange mesons, i.e., not only
to the $D^0$ meson but also to other non-strange charm mesons, the
$D^+$ and $D^{*+}$. Thus, in evaluating the transverse momentum
distribution of the $D^0$ meson in Fig.
\ref{DDvpTdistribution}(a), the feed-down contributions from two
charm-strange mesons, i.e., the $D_{s1}(2536)$ and
$D_{s2}^*(2573)$, have also been taken into account, as summarized
in Eq. (\ref{DDvfeeddown}) \cite{ParticleDataGroup:2024cfk,
Lisovyi:2015uqa}. Nevertheless, only 37.0$\%$ among the total
decay of the $D_{s2}^*(2573)$ is taken into account for the
production of the $D^0$, as the decay mode of the $D_{s2}^*(2573)$
has not been known completely \cite{ParticleDataGroup:2024cfk}.
{\allowdisplaybreaks
\begin{eqnarray}
&& \frac{dN_{D^0}^{all}}{dp_T}=\frac{dN_{D^0}}{dp_T}+
\frac{dN_{D^{*0}}}{dp_T}+0.677\frac{dN_{D^{*+}}}{dp_T}
\nonumber \\
&&\qquad\quad+(0.480\times0.677+0.400)\frac{dN_{D_{s1}(2536)}}
{dp_T} \nonumber \\
&& \qquad\quad+0.370\frac{dN_{D_{s2}^*(2573)}}{dp_T}+0.609
\frac{dN_c^{Frag}}{dp_T}, \nonumber \\
&& \frac{dN_{D^+}^{all}}{dp_T}=
\frac{dN_{D^+}}{dp_T}+0.323\frac{dN_{D^{*+}}}{dp_T},
\nonumber \\
&& \qquad\quad+(0.480\times0.323+0.120)\frac{dN_{D_{s1}(2536)}}
{dp_T} \nonumber \\
&& \qquad\quad+0.240\frac{dN_c^{Frag}}{dp_T},
\nonumber \\
&& \frac{dN_{D^{*0}}^{all}}{dp_T}=
\frac{dN_{D^{*0}}}{dp_T}+0.360\frac{dN_{D_{s1}(2536)}}{dp_T}
\nonumber \\
&& \qquad\quad~+0.231 \frac{dN_c^{Frag}}{dp_T}. \nonumber \\
&& \frac{dN_{D^{*+}}^{all}}{dp_T}=
\frac{dN_{D^{*+}}}{dp_T}+0.480\frac{dN_{D_{s1}(2536)}}{dp_T}
\nonumber \\
&& \qquad\quad~+0.243 \frac{dN_c^{Frag}}{dp_T}.
\label{DDvfeeddown}
\end{eqnarray} }
We also show in Fig. \ref{DDvpTdistribution} the transverse
momentum distributions of $D^+$, $D^{*0}$, and $D^{*+}$ mesons
produced by both coalescence and fragmentation, together with the
total transverse momentum distribution of those mesons obtained
after considering the feed-down contributions to the $D^+$,
$D^{*0}$, and $D^{*+}$, Eq. (\ref{DDvfeeddown}). The measurements
of $D^+$ and $D^{*+}$ transverse momentum distributions by ALICE
Collaboration at 0-10$\%$ centrality in Pb-Pb collisions at
$\sqrt{s_{NN}}=5.02$ TeV \cite{ALICE:2021rxa} have also been shown
in Fig. \ref{DDvpTdistribution}(b) and Fig.
\ref{DDvpTdistribution}(d) for comparison. The transverse momentum
distribution of the $D^{*0}$ in Fig. \ref{DDvpTdistribution} is
expected to be helpful when it is measured at the LHC in the
future.

The theoretical evaluations of the transverse momentum
distributions of both the $D^+$ and $D^{*+}$ are found to be lower
than the experimental measurements by ALICE Collaboration
\cite{ALICE:2021rxa}, as shown in Fig. \ref{DDvpTdistribution}.
Considering that the production of $D^+$ and $D^{*+}$ mesons by
coalescence is dominant up to intermediate transverse momenta,
i.e., about 4 and 5 GeV, as shown in Fig.
\ref{DDvpTdistribution}(b) and Fig. \ref{DDvpTdistribution}(d),
one can see that significant contributions from heavy charmed
mesons, which are also produced by coalescence, to the production
of the $D^+$ and $D^{*+}$ are missing. These must be related to
the contributions from other charmed mesons which have not been
taken into account in this work because their decay modes are not
known precisely.

Here, the transverse momentum distribution of the daughter meson,
i.e., the $D$ is assumed to be the same as those of heavier
mesons, the $D^*$, $D_{s1}(2536)$ and $D_{s2}^*(2573)$ after the
decay since the $D$ meson is much heavier compared to the other
daughter particles, e.g., the pion in the decay process.

\begin{figure}[!t]
\begin{center}
\includegraphics[width=0.50\textwidth]{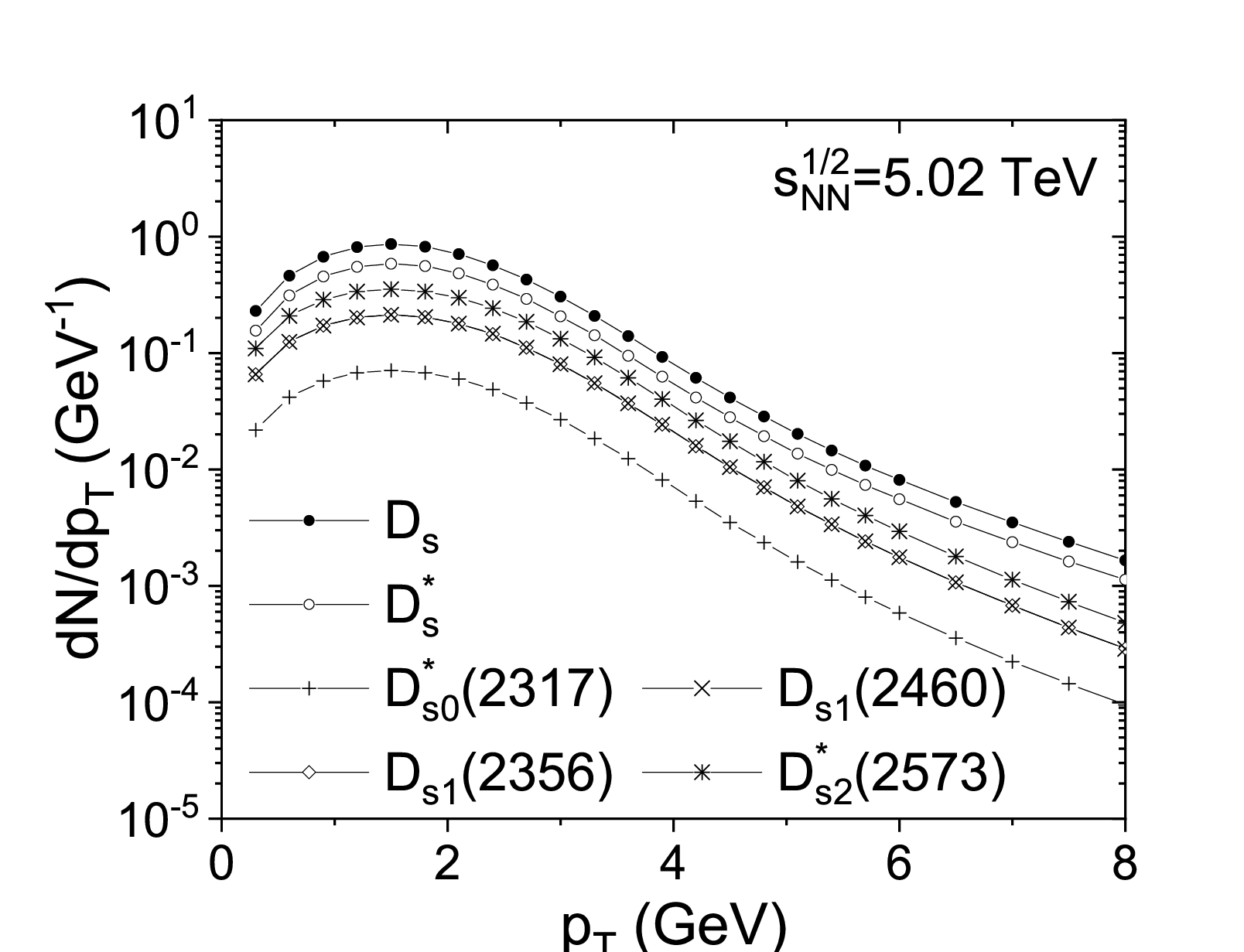}
\end{center}
\caption{Transverse momentum distributions, $dN/dp_T$ of the
$D_s$, $D_s^*$, $D_{s0}^*(2317)$, $D_{s1}(2460)$, $D_{s1}(2536)$,
and $D_{s2}^*(2573)$ at mid-rapidity at LHC $\sqrt{s_{NN}}=5.02$
TeV energy. The feed-down contributions from heavier charm-strange
mesons, i.e., $D_s^*$, $D_{s0}^*(2317)$, and $D_{s1}(2460)$ mesons
to the $D_s$ and $D_s^*$, Eq. (\ref{DsDsvfeeddown}) as well as the
production of the $D_s$ and $D_s^*$ by fragmentation from the
remaining charm quarks meson have been taken into account.}
\label{charmstrangepTdistribution}
\end{figure}
Then, we show in Fig. \ref{charmstrangepTdistribution} the
transverse momentum distributions of charm-strange mesons,
$dN/dp_T$ of the $D_s$, $D_s^*$, $D_{s0}^*(2317)$, $D_{s1}(2460)$,
$D_{s1}(2536)$, and $D_{s2}^*(2573)$ at mid-rapidity at LHC
$\sqrt{s_{NN}}=5.02$ TeV energy. In Fig.
\ref{charmstrangepTdistribution} the production of the $D_s$ and
$D_s^*$ by fragmentation from the remaining charm quarks as well
as all the following feed-down contributions from heavy
charm-strange mesons, i.e., $D_s^*$, $D_{s0}^*(2317)$, and
$D_{s1}(2460)$ mesons to the $D_s$ and $D_s^*$, Eq.
(\ref{DsDsvfeeddown}) have been taken into account
\cite{ParticleDataGroup:2024cfk, Lisovyi:2015uqa}.

\begin{eqnarray}
&& \frac{dN_{D_s}^{all}}{dp_T}= \frac{dN_{D_s}}{dp_T}
+\frac{dN_{D_s^*}}{dp_T}+\frac{dN_{D_{s0}^*(2317)}}{dp_T}
\nonumber \\
&& \qquad\quad~+0.820~\frac{dN_{D_{s1}(2460)}}{dp_T}
+0.0802\frac{dN_c^{Frag}}{dp_T}, \nonumber \\
&& \frac{dN_{D_s^*}^{all}}{dp_T}=\frac{dN_{D_s^*}}{dp_T}
+0.170\frac{dN_{D_{s0}^*(2317)}}{dp_T}
\nonumber \\
&& \qquad\quad~+(0.560+0.037\times0.170)~\frac{dN_{D_{s1}(2460)}}
{dp_T} \nonumber \\
&& \qquad\quad~+0.0548\frac{dN_c^{Frag}}{dp_T}.
\label{DsDsvfeeddown}
\end{eqnarray}

Here also, the transverse momentum distribution of the $D_s$ or
$D_s^*$ is taken to be the same as that of heavier charm-strange
mesons, the $D_{s0}^*(2317)$ and $D_{s1}(2460)$ after the decay
due to their heavier mass compared to the other daughter particles
in the decay process. It should be noted that the feed-down
contributions from the $D_{s1}(2460)$ to the $D_s$ and $D_s^*$
meson are partially taken into account; to be exact, only 82.0$\%$
of the total decay modes of $D_{s1}(2460)$ are considered, as the
decay mode of the $D_{s1}(2460)$ is not completely known
\cite{ParticleDataGroup:2024cfk}.

As shown in Fig. \ref{charmstrangepTdistribution}, the transverse
momentum distribution of the $D_{s2}^*(2573)$ is found to be
higher than that of the $D_{s1}(2460)$ while the transverse
momentum distribution of the $D_{s1}(2536)$ is similar to that of
the $D_{s1}(2460)$, implying that the contributions from
$D_{s1}(2536)$ and $D_{s2}^*(2573)$ mesons cannot be negligible
compared to those from other charm-strange mesons, the $D_s^*$,
$D_{s0}^*(2317)$ and $D_{s1}(2460)$. The similar behavior of
transverse momentum distributions of the $D_{s1}(2536)$ and
$D_{s1}(2460)$ is rather expected, as they share the same kind and
number of constituents, as well as the same spin degeneracy and
$p$-wave internal structure, shown in Eq. (\ref{CoalTransCSp}).

The slight differences in their transverse momentum distributions
are only originated from the Lorentz transformation of transverse
momenta between the fireball frame and the center of mass frame
for mesons of slightly different masses, the $D_{s1}(2536)$ and
$D_{s1}(2460)$.

\begin{figure}[!t]
\begin{center}
\includegraphics[width=0.50\textwidth]{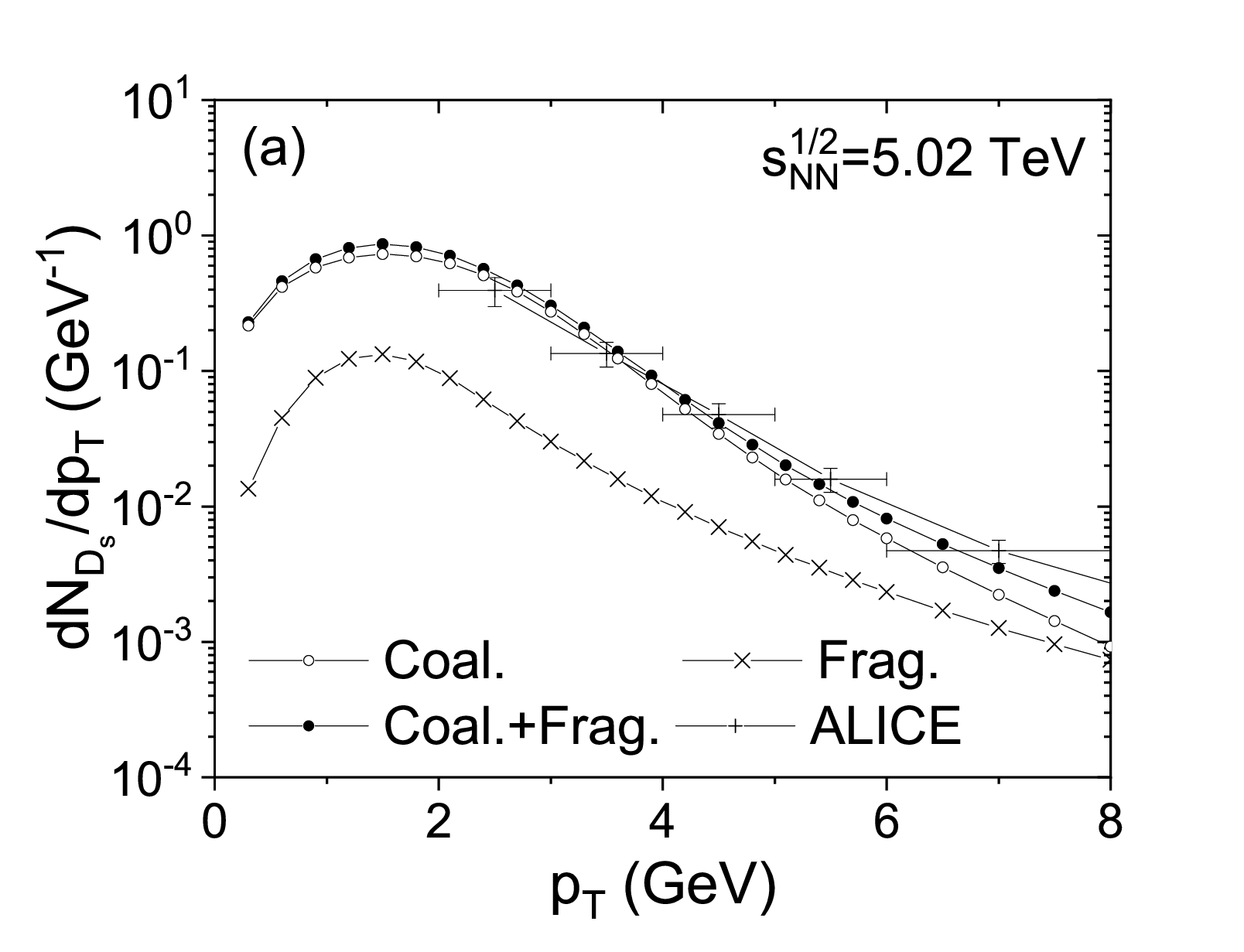}
\includegraphics[width=0.50\textwidth]{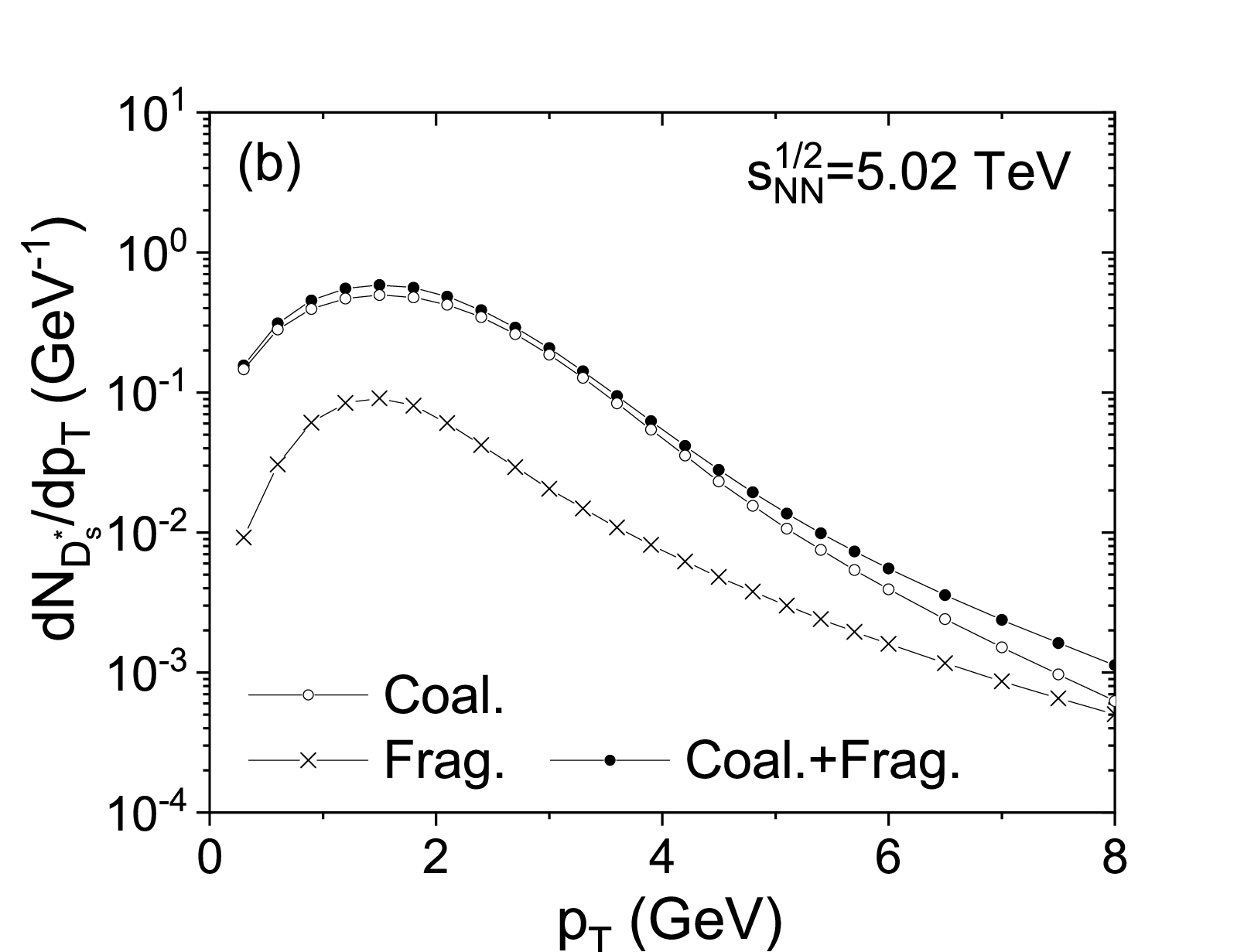}
\end{center}
\caption{(a) Transverse momentum distribution of the $D_s$ meson,
$dN_{D_s}/dp_T$ at mid-rapidity at LHC $\sqrt{s_{NN}}=5.02$ TeV
together with that of the $D_s$ produced purely by coalescence and
also by fragmentation, respectively. The measurement of the $D_s$
at the same energy by ALICE Collaboration \cite{ALICE:2021kfc} is
also shown for comparison. (b) Transverse momentum distribution of
the $D_s^*$ meson, $dN_{D_s^*}/dp_T$ at mid-rapidity together with
that of the $D_s^*$ produced purely by coalescence and also by
fragmentation, respectively. } \label{DsDsvpTdistribution}
\end{figure}

We show in Fig. \ref{DsDsvpTdistribution}(a) in more detail the
transverse momentum distribution of the $D_s$ meson,
$dN_{D_s}/dp_T$ at mid-rapidity, together with that of the $D_s$
produced purely by coalescence and also by fragmentation,
respectively. The measurement of the $D_s$ at the same energy by
ALICE Collaboration \cite{ALICE:2021kfc} is also shown for
comparison. We see in Fig. \ref{DsDsvpTdistribution}(a) that the
transverse momentum distribution of the $D_s$ meson agrees
reasonably well with the measurement of the $D_s$ meson transverse
momentum distribution by ALICE Collaboration \cite{ALICE:2021kfc}.
As also seen in Fig. \ref{DsDsvpTdistribution}(a), the $D_s$ meson
is dominantly produced by coalescence between charm and
anti-strange quarks at low and intermediate transverse momentum
regions, including feed-down contributions from heavy
charm-strange mesons while the production of $D_s$ mesons by
fragmentation is rather small compared to that of $D^0$ mesons.


\begin{widetext}

\begin{table}[!t]
\caption{Yields of charm-strange mesons, the $D_s$, $D_s^*$,
$D_{s0}^*(2317)$, and $D_{s1}(2460)$ as well as those of
non-strange charm mesons, the $D^0$, $D^+$, and $D^{*+}$ at
mid-rapidity obtained by integrating the transverse momentum
distributions shown in Figs. \ref{DDvpTdistribution} and
\ref{charmstrangepTdistribution} over all transverse momenta. Also
listed are yields of the $D_s$, $D^0$, $D^+$, and $D^{*+}$
experimentally measured at mid-rapidity, $|y|<0.5$ in 0-10$\%$
centrality class at LHC in $\sqrt{s_{NN}}=5.02$ TeV Pb+Pb
collisions by ALICE Collaboration \cite{ALICE:2021kfc,
ALICE:2021rxa}. Yields of the $D_s$, $D^0$, $D^+$, and $D^{*+}$
meson evaluated in the Statistical Hadronization Model with charm
(SHMc) \cite{Andronic:2021erx} are also presented for comparison.}
\label{yields}
\begin{center}
\begin{tabular}{c|c|c|c}
\hline \hline
Yields & Coal.  & ALICE  & SHMc  \\
\hline $D_s$ & ~1.96~ &
~1.89$\pm0.07$(stat)$^{+0.13}_{-0.16}$(syst)$^{+0.36}_{-0.55}$(extr)$\pm0.07$(BR)~
\cite{ALICE:2021kfc} & ~2.22$\pm$0.38~ \cite{Andronic:2021erx} \\
$D_s^*$ & ~1.33~ &  &  \\
$D_{s0}^*(2317)$ & ~0.167~ &  &   \\
$D_{s1}(2460)$ & ~0.502~ &  & \\
$D_{s1}(2536)$ & ~0.501~ &  &   \\
$D_{s2}^*(2573)$ & ~0.835~ &  & \\
$D^{0}$ & ~6.76~ & ~6.819$\pm0.457$(stat)$^{+0.912}_{-0.936}$(syst)$\pm0.054$(BR)~
\cite{ALICE:2021rxa} & ~6.42$\pm$1.07~ \cite{Andronic:2021erx}  \\
$D^{+}$ & ~2.26~ & ~3.041$\pm0.073$(stat)$^{+0.154}_{-0.155}$(syst)$\pm0.052$(BR)$^{+0.352}_{-0.618}$(extrap.)~
\cite{ALICE:2021rxa} & ~2.84$\pm$0.47~ \cite{Andronic:2021erx} \\
$D^{*0}$ & ~2.98~ &   &  \\
$D^{*+}$ & ~3.08~ &  ~3.803$\pm0.037$(stat)$^{+0.084}_{-0.085}$(syst)$\pm0.041$(BR)$^{+0.854}_{-0.175}$(extrap.)~
\cite{ALICE:2021rxa} & ~2.52$\pm$0.42~ \cite{Andronic:2021erx} \\
\hline \hline
\end{tabular}
\end{center}
\end{table}

\end{widetext}

We also show in Fig. \ref{DsDsvpTdistribution}(b) the transverse
momentum distribution of the $D_s^*$ meson, $dN_{D_s^*}/dp_T$ at
mid-rapidity, together with that of the $D_s^*$ produced purely by
coalescence and also by fragmentation, respectively. Similarly to
the production of the $D_s$ meson, the contribution to the
production of the $D_s^*$ by fragmentation does not prevail in the
transverse momentum range up to about $p_T=8.0$ GeV. Here, the
feed-down contributions from $D_{s0}^*(2317)$, and $D_{s1}(2460)$
mesons to the $D_s^*$, Eq. (\ref{DsDsvfeeddown}) have been taken
into account as well.

As mentioned before, the feed-down contributions from the other
two heavy charm-strange mesons, the $D_{s1}(2536)$ and
$D_{s2}^*(2573)$, to the $D_s^*$ have not been included here, as
no explicit decay modes from the above two mesons to the $D_s^*$
have been known. Nevertheless, there still exist slight
possibilities, especially for the $D_{s1}(2536)$ to decay to the
$D_s$ and $D_s^*$ \cite{ParticleDataGroup:2024cfk}, and therefore
it should be noted that corrections of the yields or transverse
momentum distributions of the $D_s$ and $D_s^*$ based on the
feed-down contributions, Eq. (\ref{DsDsvfeeddown}) discussed here
may be necessary in the future.

The yields of the charmed mesons can be obtained by integrating
the transverse momentum distributions shown in Figs.
\ref{DDvpTdistribution} and \ref{charmstrangepTdistribution} over
all transverse momenta. We show in Table \ref{yields} the yields
of charm-strange mesons, the $D_s$, $D_s^*$, $D_{s0}^*(2317)$, and
$D_{s1}(2460)$ as well as those of non-strange charm mesons, the
$D^0$, $D^+$, $D^{*0}$ and $D^{*+}$ at mid-rapidity. Also listed
in Table \ref{yields} are yields of the $D_s$, $D^0$, $D^+$, and
$D^{*+}$ experimentally measured at mid-rapidity, $|y|<0.5$ in
0-10$\%$ centrality class at LHC in $\sqrt{s_{NN}}=5.02$ TeV Pb+Pb
collisions by ALICE Collaboration \cite{ALICE:2021kfc,
ALICE:2021rxa}. The yields of the $D_s$, $D^0$, $D^+$, and
$D^{*+}$ meson evaluated in the Statistical Hadronization Model
with charm (SHMc) \cite{Andronic:2021erx} are also presented for
comparison.

The feed-down contributions from heavier mesons for the $D_s$,
$D^0$, $D^+$, and $D^{*+}$, Eqs. (\ref{DDvfeeddown}) and
(\ref{DsDsvfeeddown}) are taken into account in calculating the
yield listed in Table \ref{yields}. The similar transverse
momentum distributions of the $D_{s1}(2536)$ and $D_{s1}(2460)$
discussed earlier result in nearly the same yields for those
mesons, as shown in Table \ref{yields}.

The yield of the $D^0$ meson agrees more reasonably with that of
the measurement by ALICE Collaboration \cite{ALICE:2021rxa} and
the yield of the $D_s$ meson becomes closer to that of the
measurement also by ALICE Collaboration \cite{ALICE:2021kfc},
compared to the previous results on charm-strange meson
production, 6.54 and 2.18 for the $D^0$ and $D_s$ yields,
respectively \cite{Cho:2025lrc}. Taking into account the
production of additional two charm-strange mesons, $D_{s1}(2536)$
and $D_{s2}^*(2573)$ which do not decay to charm-strange mesons,
increases the fractions of charm quarks contributing to the
production of the $D^0$ meson, and eliminates at the same time
possibilities for the production of charm-strange mesons, the
$D_s$ and $D_s^*$, thereby increasing and reducing the yield of
the $D^0$ and $D_s$, respectively.

It is noticeable that the yields of heavier charm-strange mesons,
i.e., the $D_{s0}^*(2317)$, $D_{s1}(2460)$, $D_{s1}(2536)$ and
$D_{s2}^*(2573)$ are comparable to those of the $D_s$, $D_s^*$,
implying that their contributions could not be ignored in
explaining the production of the $D_s$ meson. The yields of
$D_{s0}^*(2317)$ and $D_{s1}(2460)$ mesons are responsible for
about 30$\%$ of the $D_s$ meson production.

As expected in Fig. \ref{DDvpTdistribution}, the yields of the
$D^+$ and $D^{*+}$ are evaluated to be smaller than the
measurements by ALICE Collaboration \cite{ALICE:2021rxa}, as shown
in Table. \ref{yields}. The yields of the $D^+$ and $D^{*+}$ are
smaller, respectively, by 26$\%$ and 19$\%$ than the measurements
of those at the LHC. Similarly, the $D^{*0}$ yield in Table
\ref{yields} is expected to be also smaller than the measurement
of that at the LHC when it is measured in the future.


\begin{widetext}

\begin{figure}[!b]
\begin{center}
\includegraphics[width=0.495\textwidth]{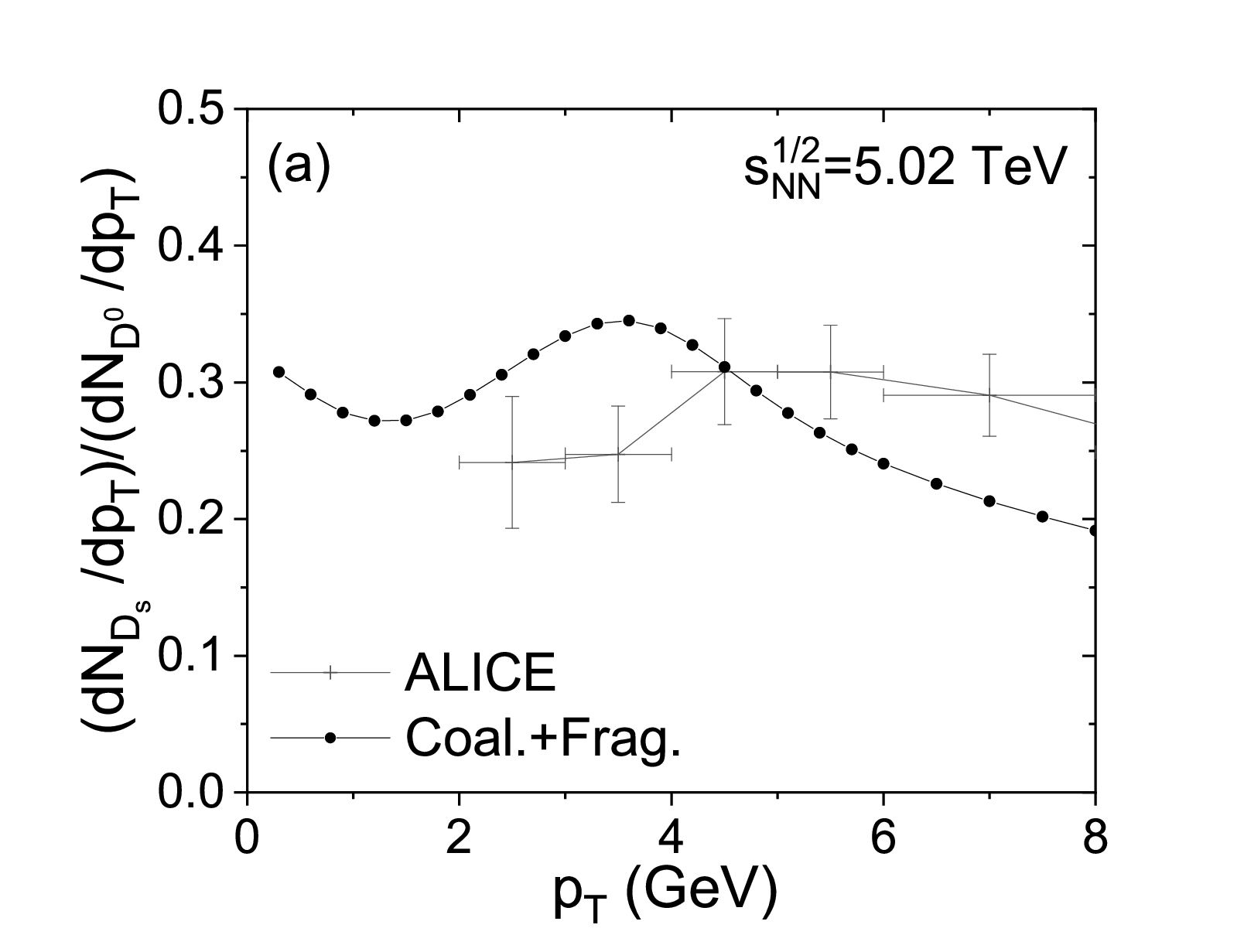}
\includegraphics[width=0.495\textwidth]{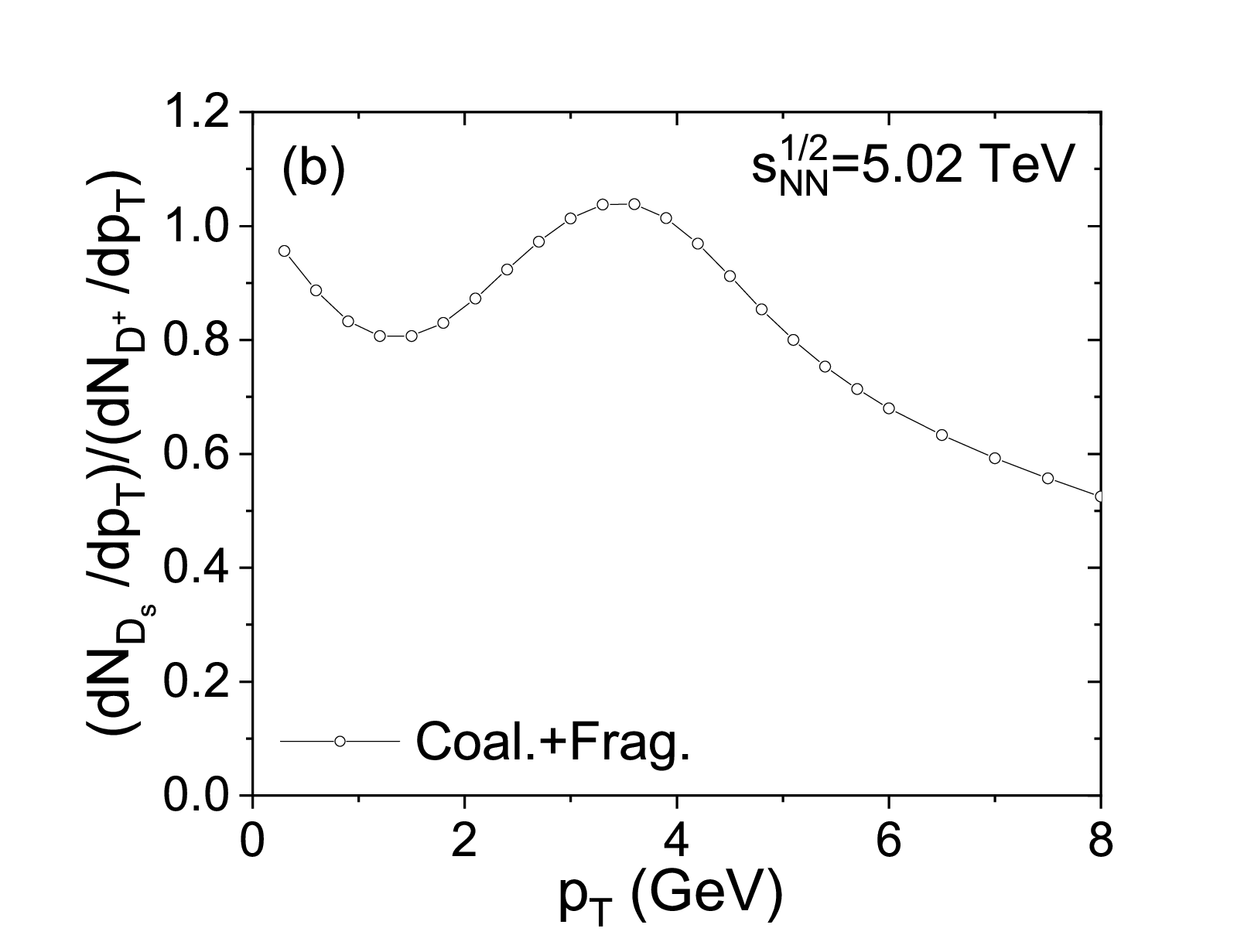}
\includegraphics[width=0.495\textwidth]{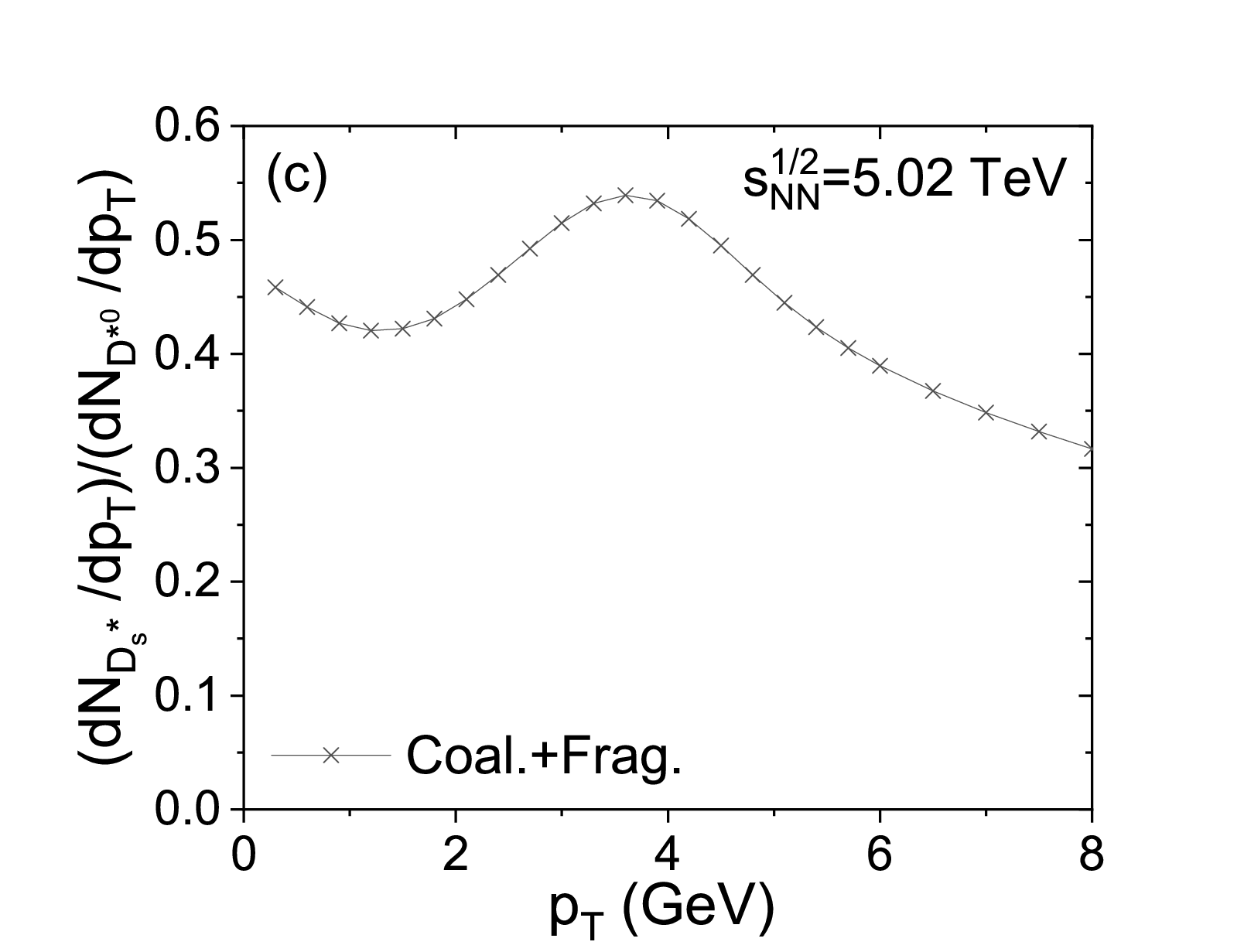}
\includegraphics[width=0.495\textwidth]{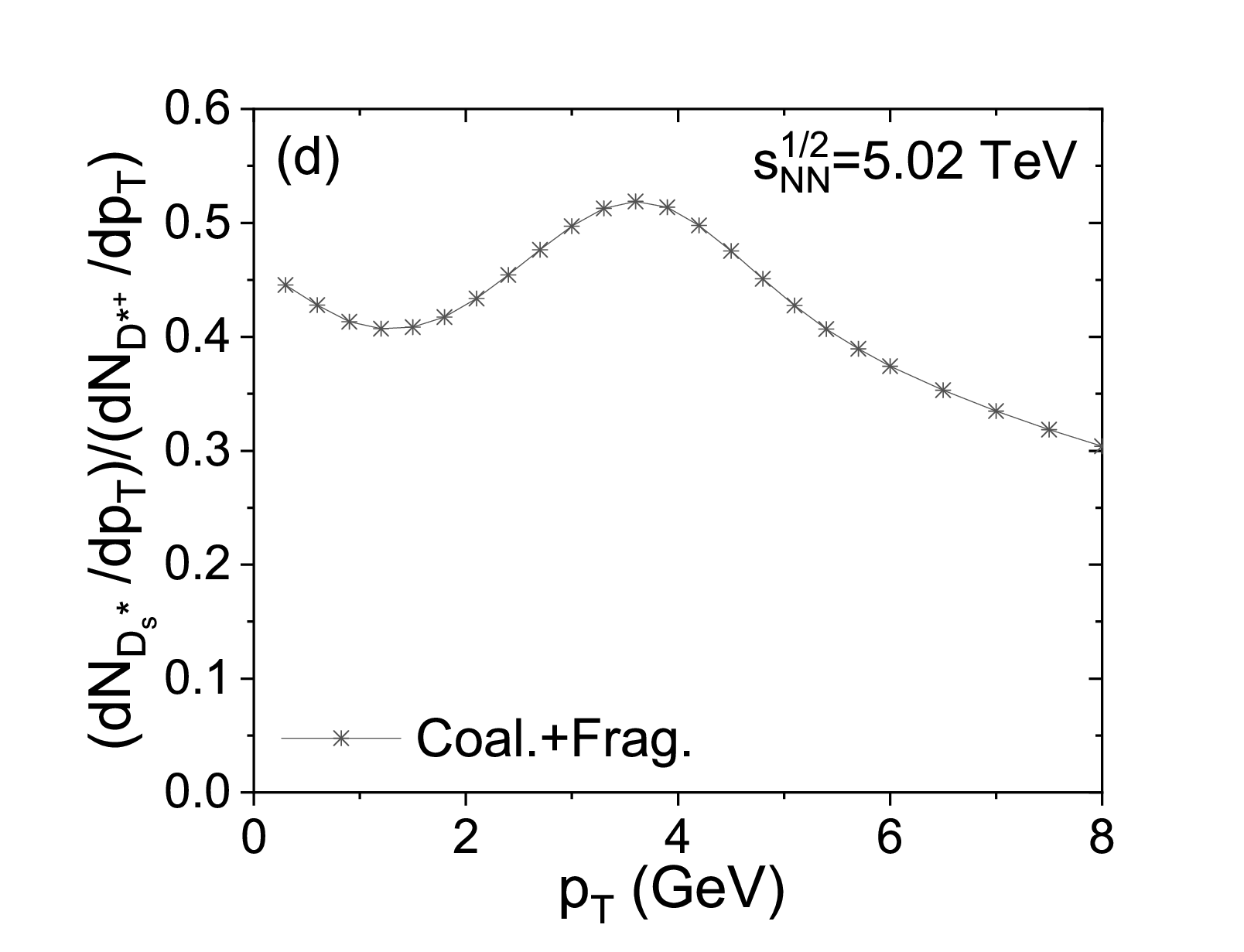}
\end{center}
\caption{Transverse momentum distribution ratios between strange
and non-strange charmed mesons, $D_s/D^0$ (a), $D_s/D^+$ (b),
$D_s^*/D^{*0}$ (c), and $D_s^*/D^{*+}$ (d) at mid-rapidity in
$0-10\%$ centrality at $\sqrt{s_{NN}}=5.02$ TeV at LHC. Also shown
in (a) is the same transverse momentum distribution ratio between
$D_s$ and $D^0$ mesons measured by ALICE Collaboration at
$\sqrt{s_{NN}}=5.02$ TeV at LHC \cite{ALICE:2021kfc}. }
\label{DsDsvtoDDvratio}
\end{figure}

\end{widetext}

The possible explanation for the smaller yields of the $D^+$ and
$D^{*+}$ must be related to heavier hadrons that have been
observed but are not included in the discussion, as their decay
modes are not fully known. As mentioned before, we have only taken
into account only 82.0$\%$ and 37.0$\%$ of the total decay
contributions from the $D_{s1}(2460)$ and $D_{s2}^*(2573)$,
respectively in evaluating the transverse momentum distributions
and yields of the $D$, $D^*$, $D_s$ and $D_s^*$ as the other
remaining decay contributions from the $D_{s1}(2460)$ and
$D_{s2}^*(2573)$ have not been known. Moreover, there are still
many scarcely known charmed mesons with masses below 3 GeV, i.e.,
five non-strange charmed mesons, $D_0^*(2300)$, $D_1(2420)$,
$D_1(2430)$, $D_2^*(2460)$ and $D_3^*(2750)$, and two strange
mesons, $D_{s1}^*(2700)$ and $D_{s3}^*(2860)$, which are expected
to decay mostly to light charmed mesons, $D$ and $D^*$ mesons. As
the transverse momentum distributions of $D_{s1}(2536)$ and
$D_{s2}^*(2573)$ mesons are not negligible compared to other
charm-strange mesons, it is expected that the inclusion of the
above eight mesons helps in understanding the production of $D^+$
and $D^{*+}$ mesons in more detail. It is also possible that there
might be other heavier charmed hadrons that have not yet been
discovered but might have already decayed to the $D^+$ and
$D^{*+}$.

\section{Transverse momentum distribution ratios between charmed mesons}

\begin{widetext}

\begin{center}
\begin{table}[!h]
\caption{Possible decay modes and decay products of the $D_s$,
$D^+$, $D_{s0}^*(2317)$, $D_s^*$, $D_{s1}(2536)$, and
$D_{s2}^*(2573)$ together with their branching ratios
\cite{ParticleDataGroup:2024cfk}.} \label{decaymodes}
\begin{tabular}{c|c|c}
\hline \hline
Decay Process & Branching & Decay  \\
 & Ratio ($\%$) & Products  \\
\hline $D_s\to\phi\pi^+\to K^+K^-\pi^+$ & $4.5;50.1$ & $K^+K^-\pi^+$ \\
$D^+\to K^+K^-(\phi)\pi^+$ & $2.69\times 10^{-3}$ &  \\
\hline $D_s^*\to D_s\pi^0\to\phi\pi^+\pi^0\to K^+K^-\pi^+\pi^0$ & $5.7;4.5;50.1$ & $K^+K^-\pi^+\pi^0$ \\
$D^{*+}\to D^+\pi^0\to K^+K^-(\phi)\pi^+\pi^0$ & $30.7;2.69\times 10^{-3}$ &  \\
\hline
$D_{s0}^*(2317) \to D_s\pi^0$ & $100$ & $D_s\pi^0$ \\
$D_s^{*}\to D_s\pi^0$ & $5.77$ &  \\
\hline
$D_{s1}(2536)\to D^0K^+\to K^-\pi^+K^+$ & $<12;3.9$ & $K^+K^-\pi^+$ \\
$D_{s2}^*(2573)\to D^0K^+\to K^-\pi^+K^+$ & $37;3.9$ &  \\
$D^{*+}\to D^0\pi^+\to K^+K^-\pi^+$ & $67.7;4.07\times 10^{-3}$ &  \\
\hline
$D_{s1}(2536)\to D^0K^+\to K^-\pi^+\pi^0K^+$ & $<12;14.4$ & $K^+K^-\pi^+\pi^0$ \\
$D_{s2}^*(2573)\to D^0K^+\to K^-\pi^+\pi^0K^+$ & $37;14.4$ &  \\
$D^{*+}\to D^+\pi^0\to K^+K^-\pi^+\pi^0$ & $30.7;9.61\times 10^{-3}$ &  \\
\hline \hline
\end{tabular}
\end{table}
\end{center}

\end{widetext}

\subsection{Transverse momentum distribution ratio between
strange and non-strange charmed mesons}

We investigate in more detail the production of charmed mesons by
comparing the transverse momentum distributions of charm-strange
mesons to those of non-strange charmed mesons. As all the
transverse momentum distributions of non-strange charmed mesons,
i.e., the $D^0$, $D^+$, $D^{*0}$, and $D^{*+}$, are available, it
is possible to take the various transverse momentum distribution
ratios between charm-strange and non-strange charmed mesons,
enabling us to look into the role of the strange quark in the
hadronization process of charm-strange mesons from a more diverse
point of view. Here we not only compare the production of the
$D_s$ to that of the $D^0$ and $D^+$ but also compare the
production of the $D_s^*$ to that of $D^{*0}$ and $D^{*+}$ mesons.

Recently, the ALICE Collaboration has measured the transverse
momentum distribution ratio between the $D_s$ and $D^+$ from the
measurement of the invariant mass of $K^+K^-\pi^+$ utilizing the
common decay modes of the $D_s$ and $D^+$, $D_s\to \phi\pi^+\to
K^+K^-\pi^+$ and $D^+\to K^+K^-\pi^+$, respectively
\cite{Chinu:2026ouk}.

Similarly, considering the common decay modes between the $D_s^*$
and $D^{*+}$, i.e., $D_s^*\to D_s\pi^0\to\phi\pi^+\pi^0\to
K^+K^-\pi^+\pi^0$ and either $D^{*+}\to D^+\pi^0\to
K^+K^-\pi^+\pi^0$ or $D^{*+}\to D^0\pi^+\to K^+K^-\pi^0\pi^+$,
respectively, the transverse momentum distribution ratio between
the $D_s^*$ and $D^{*+}$ would be also helpful in investigating
the strangeness production in heavy ion collisions more
explicitly. Additionally, we consider the transverse momentum
distribution ratio between the $D_s^*$ and $D^{*0}$, though the
$D_s^*$ and $D^{*0}$ do not share common decay modes as they have
different electric charges.

We take the transverse momentum distribution ratio of the $D_s$
shown in Fig. \ref{DsDsvpTdistribution}(a) to the $D^0$ in Fig.
\ref{DDvpTdistribution}(a), and that of the $D_s$ to the $D^+$ in
Fig. \ref{DDvpTdistribution}(b). We also calculate the transverse
momentum distribution ratio of the $D_s^*$ shown in Fig.
\ref{DsDsvpTdistribution}(b) to the $D^{*0}$ in Fig.
\ref{DDvpTdistribution}(c), and that of the $D_s^*$ to the
$D^{*+}$ in Fig. \ref{DDvpTdistribution}(d), and show results in
Fig. \ref{DsDsvtoDDvratio}. Also shown in Fig.
\ref{DsDsvtoDDvratio}(a) is the transverse momentum distribution
ratio between the $D_s$ and $D^0$ meson measured by ALICE
Collaboration at $\sqrt{s_{NN}}=5.02$ TeV at the LHC
\cite{ALICE:2021kfc}.

The transverse momentum distribution ratio between $D_s$ and $D^0$
mesons is evaluated to be higher at low transverse momenta, and
lower at high transverse momenta compared to the measurement by
ALICE Collaboration \cite{ALICE:2021kfc}. The peak of the ratio is
found to be located at about $p_T=4$ GeV, also lower than that of
the measurement, at about $p_T=5$ GeV, as shown in Fig.
\ref{DsDsvtoDDvratio}(a). The ratio lies between 0.2 and 0.4,
within the yield ratio between $D_s$ and $D^0$ mesons,
$1.96/6.84\approx 0.287$, Table \ref{yields}.

All the transverse momentum distribution ratios between the $D_s$
or $D_s^*$ and the $D^0$ or $D^{*0}$ behave similarly as shown in
Fig. \ref{DsDsvtoDDvratio}, exhibiting an enhancement in the
intermediate transverse momentum regions with its maximum at about
$p_T=4$ GeV. The transverse momentum distribution ratio between
the $D_s$ and $D^+$ is found to be about three times higher than
that between the $D_s$ and $D^0$, reflecting the yield difference
of the same amount between the $D^0$ and $D^+$, Table
\ref{yields}. Similarly, the transverse momentum distribution
ratio between the $D_s^*$ and $D^{*+}$ and that between the
$D_s^*$ and $D^{*0}$ are evaluated to be almost the same, also
representing almost an equal amount of yield between the $D^{*+}$
and $D^{*0}$, Table \ref{yields}.

\subsection{Transverse momentum distribution ratio between
heavier charm-strange mesons and $D_s^*$ or $D^*$}

\begin{widetext}

\begin{figure}[!h]
\begin{center}
\includegraphics[width=0.49\textwidth]{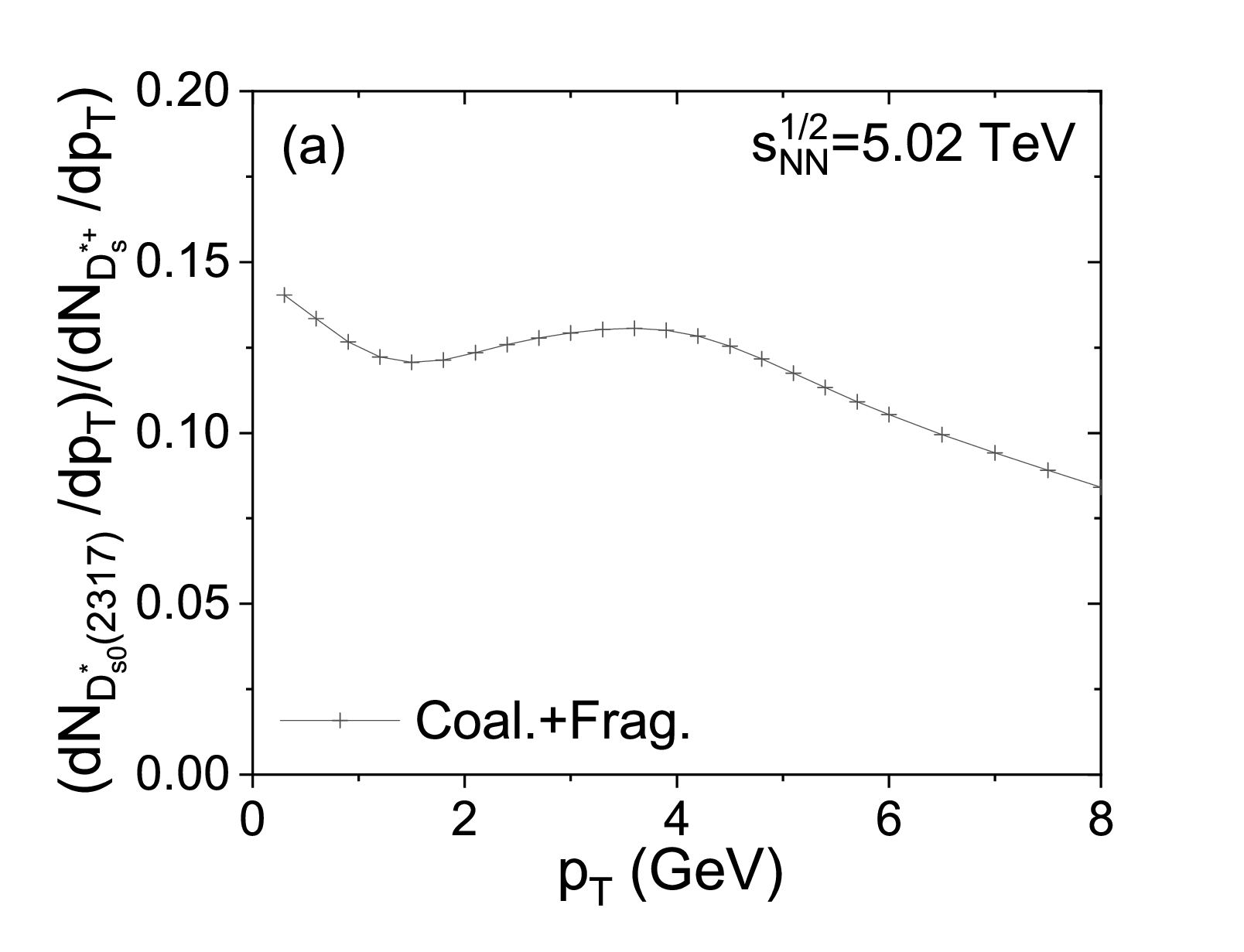}
\includegraphics[width=0.49\textwidth]{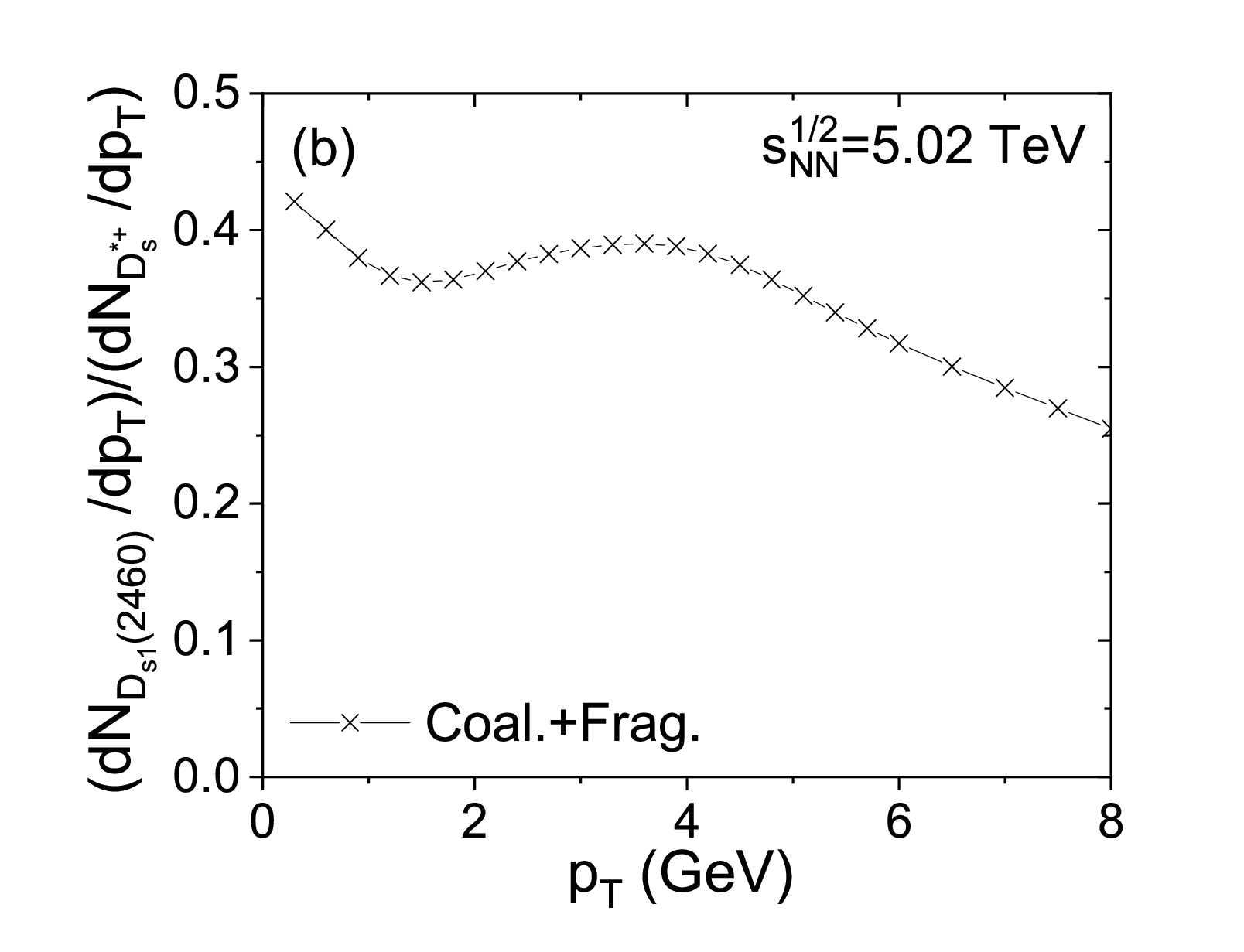}
\includegraphics[width=0.49\textwidth]{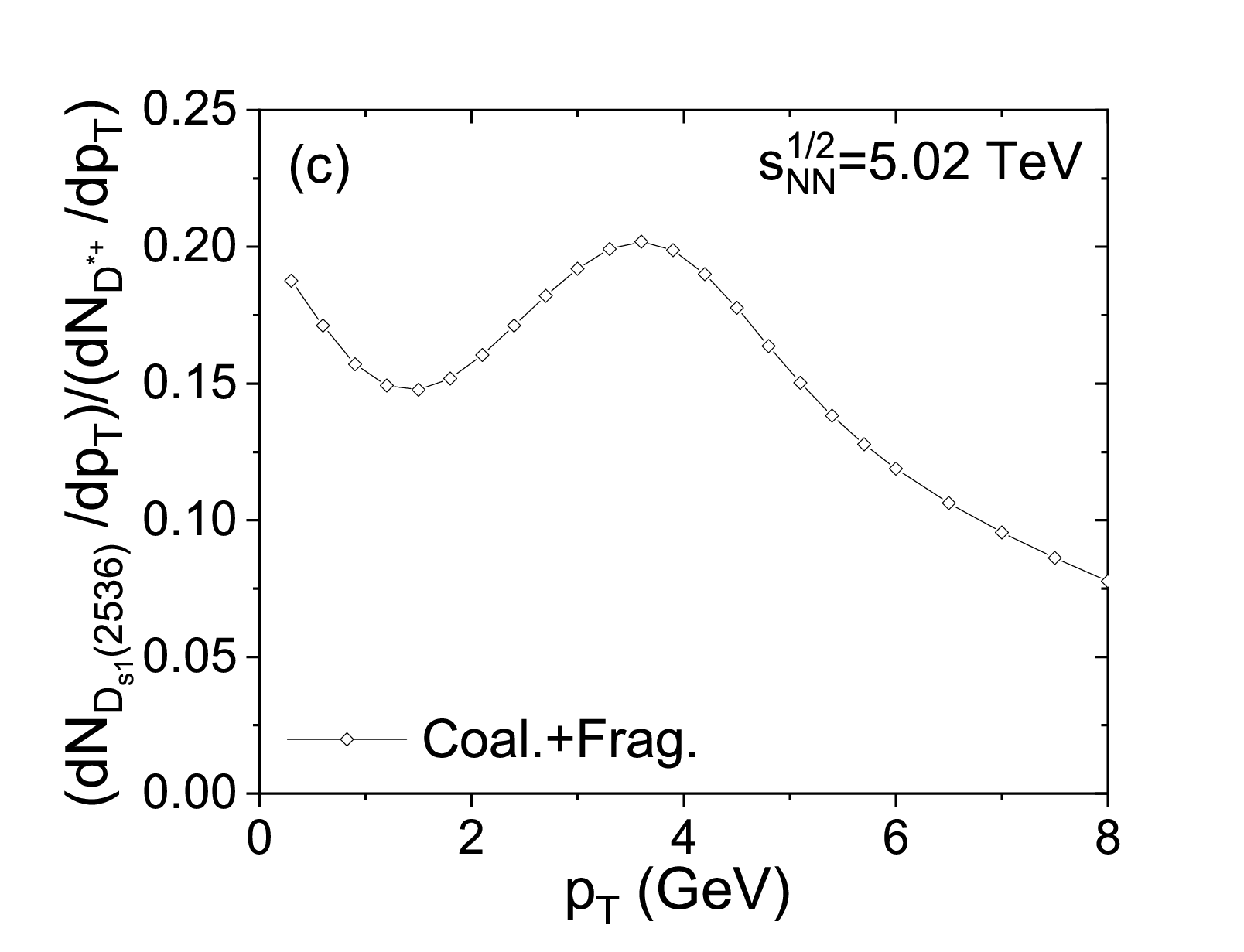}
\includegraphics[width=0.49\textwidth]{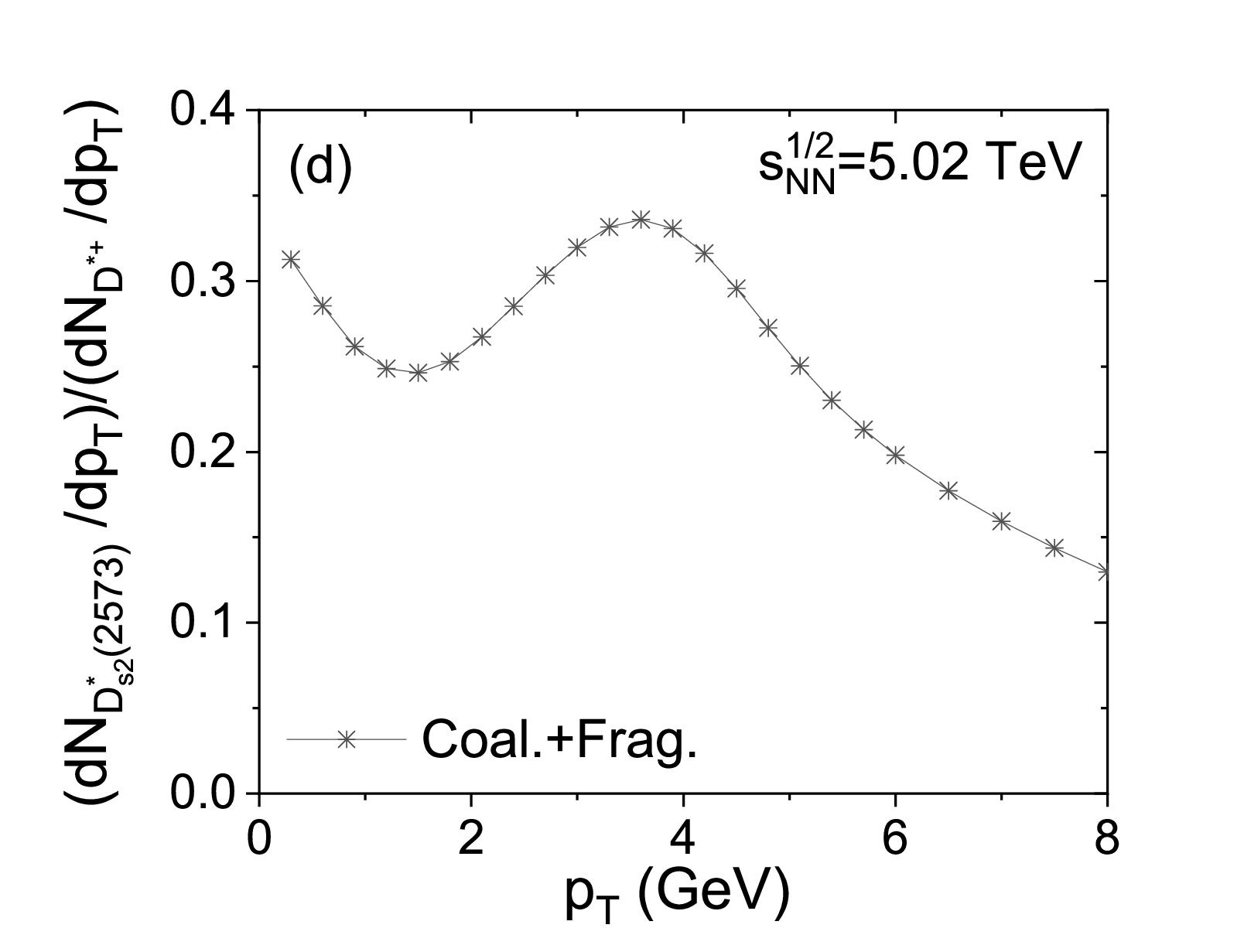}
\end{center}
\caption{Transverse momentum distribution ratios between the
$D_{s0}^*(2317)$ and $D_s^*$,
$(dN_{D_{s0}^*(2317)}/dp_T)/(dN_{D_s^*}/dp_T)$ (a), between the
$D_{s1}(2460)$ and $D_s^*$,
$(dN_{D_{s1}(2460)}/dp_T)/(dN_{D_s^*}/dp_T)$ (b), between the
$D_{s1}(2536)$ and $D^{*+}$,
$(dN_{D_{s1}(2536)}/dp_T)/(dN_{D^{*+}}/dp_T)$ (c) and between the
$D_{s2}^*(2573)$ and $D^{*+}$ (d),
$(dN_{D_{s2}^*(2573)}/dp_T)/(dN_{D^{*+}}/dp_T)$ at mid-rapidity in
central collisions at the LHC $\sqrt{s_{NN}}=5.02$ TeV energy,
respectively.} \label{DsheaviertoDsvDv}
\end{figure}

\end{widetext}

We present here the transverse momentum distribution ratios
between the heavier charm-strange mesons, i.e., the
$D_{s0}^*(2317)$, $D_{s1}(2460)$, $D_{s1}(2536)$, and
$D_{s2}^*(2573)$, and the $D_s^*$ or $D^*$. As already mentioned
in the previous subsection, charm-strange and charm mesons, i.e.,
the $D_s$ and $D$ or the $D_s^*$ and $D^*$, share common decay
modes. Similarly, the common decay modes between the heavier
charm-strange mesons and the $D_s^*$ or $D^*$ can be found, and
thereby the heavier charm-strange mesons are expected to be
measured more consistently by cancelling the possible common
uncertainties when the invariant mass of their decay products is
also observed together with that of the lighter charm-strange or
charmed mesons simultaneously.

As the $D_{s0}^*(2317)$ shares a common decay mode with the
$D_s^*$, $D_{s0}^*(2317) \to D_s \pi^0~$ and $D_s^* \to D_s\pi^0$,
it would be helpful to evaluate the transverse momentum
distribution ratio between the $D_{s0}^*(2317)$ and the $D_s^*$ in
measuring the $D_{s0}^*(2317)$ in heavy ion collisions.

In addition, we also provide a comparison of the transverse
momentum distributions between the $D_{s1}(2460)$ and $D_s^*$.
Though it is not possible to find common decay products from the
$D_{s1}(2460)$ and $D_s^*$, it would be useful to compare the
production of the $D_{s1}(2460)$ to that of the $D_s^*$.

Similarly, one notes that there exist no common decay modes
between the $D_{s1}(2536)$ and $D^{*+}$, nor between the
$D_{s2}^*(2573)$ and $D^{*+}$. Nevertheless, the $D_{s1}(2536)$,
$D_{s2}^*(2573)$, and $D^{*+}$ share final decay products; the
decays of $D_{s1}(2536)$, $D_{s1}(2536)\to D^0 K^+\to K^-\pi^+
K^+$ or $K^+K^-\pi^+\pi^0$ corresponds to those of the $D^{*+}$,
$D^{*+}\to D^0\pi^+\to K^+K^-\pi^+$ or $D^{*+}\to D^+\pi^0\to
K^+K^-\pi^+\pi^0$, respectively. Similarly, the decay mode of the
$D_{s2}^*(2573)$, $D_{s2}^*(2573)\to D^0 K^+\to K^-\pi^+\pi^0K^+$
or $K^+K^-\pi^+\pi^0$ corresponds to those of the $D^{*+}$,
$D^{*+}\to D^0\pi^+\to K^+K^-\pi^+$ or $D^{*+}\to D^+\pi^0\to
K^+K^-\pi^+\pi^0$, respectively. Thus, we present here the
comparison of the transverse momentum distributions between the
$D_{s1}(2536)$ or $D_{s2}^*(2573)$ and the $D^{*+}$.

We summarize in Table \ref{decaymodes} the common decay products
of the $D_s$ and $D^+$, those of the $D_s^*$ and $D^{*+}$, those
of the $D_{s0}^*(2317)$ and $D_s^*$, those of the $D_{s1}(2536)$
and $D^{*+}$, and those of the $D_{s2}^*(2573)$ and $D^{*+}$
together with their branching ratios
\cite{ParticleDataGroup:2024cfk}, and show in Fig.
\ref{DsheaviertoDsvDv} the transverse momentum distribution ratios
between the $D_{s0}^*(2317)$ and $D_s^*$, between the
$D_{s1}(2460)$ and $D_s^*$, between the $D_{s1}(2536)$ and
$D^{*+}$ and between the $D_{s2}^*(2573)$ and $D^{*+}$ at
mid-rapidity in central collisions at the LHC $\sqrt{s_{NN}}=5.02$
TeV energy, respectively.

The transverse momentum distribution ratios between the
$D_{s0}^*(2317)$ and $D_s^*$ and that between the $D_{s1}(2460)$
and $D_s^*$ are found to be monotonically decreasing with a
shallow peak in intermediate transverse momentum regions, as shown
in Fig. \ref{DsheaviertoDsvDv}(a) and (b). Remind that the ratio
between the $D_{s0}^*(2317)$ and $D_s^*$ and that between the
$D_{s1}(2460)$ and $D_s^*$ are ratios between charm-strange
mesons, $c\bar{s}/c\bar{s}$, thereby only causing minor
differences in their production owing to different coalescence
probabilities originating from the mass differences between
charm-strange mesons.

On the other hand, the transverse momentum distribution ratios
between the $D_{s1}(2536)$ and $D^{*+}$ and between the
$D_{s2}^*(2573)$ and $D^{*+}$ are the ratio between strange and
non-strange charmed mesons, $c\bar{s}/c\bar{d}$ and thus show
similar behaviors as those ratios between the $D_s$ and $D^0$ or
that between the $D_s$ and $D^+$ shown in Fig.
\ref{DsDsvtoDDvratio}, exhibiting the peak at intermediate
transverse momentum regions. The appearance of explicit peaks in
the ratio is largely attributable to differences in the transverse
momentum distribution between a strange and light quark in the
$D_{s1}(2536)$ or $D_{s2}^*(2573)$ and $D^{*+}$, leading to
different coalescence probability for forming $D_{s1}(2536)$ or
$D_{s2}^*(2573)$ and $D^{*+}$ mesons \cite{Cho:2025lrc}.

\section{Conclusions}

We have studied charmed mesons in relativistic heavy-ion
collisions, focusing on their production from the quark-gluon
plasma by quark recombination. We have considered the production
of not only the $D$, $D^*$, $D_s$, and $D_s^*$, but also the
$D_{s0}^*(2317)$, $D_{s1}(2460)$, $D_{s1}(2536)$, and
$D_{s2}^*(2573)$, thereby investigating the contributions from
heavier charm-strange resonances to the production of the $D$ and
$D_s$ mesons at the Large Hadron Collider at $\sqrt{s_{NN}}=5.02$
TeV.

To this end, we first have obtained the new charm quark transverse
momentum distribution based on the transverse momentum
distribution of the $D^0$ meson measured at mid-rapidities in
central collisions in relativistic heavy ion collisions at
$\sqrt{s_{NN}}=5.02$ TeV by ALICE Collaboration, after taking into
account the production of the above 8 charmed mesons and 10
charmed baryons. In calculating the transverse momentum
distribution of $D^0$ mesons, we have also included the
contribution from the production of $D^0$ mesons by fragmentation
in high transverse momentum regions. Then, we have calculated the
transverse momentum distributions of 8 charmed mesons mentioned
above using the coalescence model and have also evaluated their
yields of the charmed mesons by carrying out the integration of
the transverse momentum distributions over all transverse momentum
regions.

Finally, we have compared the production of charm-strange mesons
to that of non-strange charmed mesons, i.e., the $D_s$ to $D^0$,
the $D_s$ to $D^+$, the $D_s^*$ to $D^{*0}$, and the $D_s^*$ to
$D^{*+}$. In addition, we have also compared the production of
heavier charm-strange mesons, i.e., the $D_{s0}^*(2317)$,
$D_{s1}(2460)$, $D_{s1}(2536)$, and $D_{s2}^*(2573)$, to that of
the $D_s^*$ and $D^{*+}$, presenting chances to measure the
heavier charm-strange mesons more consistently based on the
invariant masses of their common decay products.

We find that the yield of the $D_s$ meson becomes closer to the
measurement by ALICE Collaboration than in the previous results on
charm-strange meson production, which did not consider the
production of the $D_{s1}(2536)$ and $D_{s2}^*(2573)$ meson. We
observe that taking into account the production of additional two
charm-strange mesons, $D_{s1}(2536)$ and $D_{s2}^*(2573)$ which do
not decay to charm-strange mesons, increases the fractions of
charm quarks contributing to the production of the $D^0$ meson and
eliminates at the same time possibilities for the production of
charm-strange mesons, the $D_s$ and $D_s^*$, thereby increasing
and reducing the yield of the $D^0$ and $D_s$, respectively.

The yields of heavier charm-strange mesons, i.e., the
$D_{s0}^*(2317)$, $D_{s1}(2460)$, $D_{s1}(2536)$ and
$D_{s2}^*(2573)$ are evaluated to be comparable to those of the
$D_s$, $D_s^*$, indicating that their contributions should not be
ignored in understanding the production of the $D^0$ and $D_s$
meson; the yields of the $D_{s0}^*(2317)$ and $D_{s1}(2460)$ meson
are responsible for about 30$\%$ of the $D_s$ meson production.

In contrast, the yields of the $D^+$ and $D^{*+}$ are calculated
to be smaller than the measurements by ALICE Collaboration. The
yields of the $D^+$ and $D^{*+}$ are smaller, respectively, by
26$\%$ and 19$\%$ than the measurements of those at the LHC. The
plausible explanation of the smaller yields for the $D^+$ and
$D^{*+}$ must be related to heavier charmed mesons which have been
observed, but are not included in the discussion as their decay
modes have not been completely known, e.g., the $D_0^*(2300)$,
$D_1(2420)$, $D_1(2430)$, $D_2^*(2460)$, $D_3^*(2750)$,
$D_{s1}^*(2700)$ and $D_{s3}^*(2860)$ meson. Moreover, it is also
possible that there might exist other heavier charmed hadrons
which have not been discovered yet, but might have already decayed
to the $D^+$ and $D^{*+}$.

When the production of charm-strange mesons is compared to that of
non-strange charmed mesons, all the transverse momentum
distribution ratios between the $D_s$ or $D_s^*$ and the $D^0$ or
$D^{*0}$ show similar characteristics, exhibiting an enhancement
in the intermediate transverse momentum regions. The transverse
momentum distribution ratio between the $D_s^*$ and $D^{*+}$ and
that between the $D_s^*$ and $D^{*0}$ are evaluated to be almost
the same, representing almost an equal amount of yield between the
$D^{*+}$ and $D^{*0}$.

In the comparison of the production of heavier charm-strange
mesons, i.e., the $D_{s0}^*(2317)$, $D_{s1}(2460)$, $D_{s1}(2536)$
and $D_{s2}^*(2573)$ to that of the $D_s^*$ and $D^{*+}$, the
transverse momentum distribution ratios between the
$D_{s0}^*(2317)$ and $D_s^*$ and that between the $D_{s1}(2460)$
and $D_s^*$ are found to be monotonically decreasing with the
shallow peak at intermediate transverse momentum regions,
originated from mostly mass differences in coalescence
probabilities as expected from the comparison between
charm-strange mesons.

On the other hand, the transverse momentum distribution ratios
between the $D_{s1}(2536)$ and $D^{*+}$ and between the
$D_{s2}^*(2573)$ and $D^{*+}$, or the ratio between strange and
non-strange charmed mesons, show similar behaviors as those ratios
between the $D_s$ and $D^0$ or that between the $D_s$ and $D^+$,
exhibiting the peak at intermediate transverse momentum regions.
The appearance of explicit peaks in the ratio is largely
attributable to differences in the transverse momentum
distribution between a strange and light quark in the
$D_{s1}(2536)$ or $D_{s2}^*(2573)$ and $D^{*+}$, resulting in
different coalescence probabilities for forming $D_{s1}(2536)$ or
$D_{s2}^*(2573)$ and $D^{*+}$ mesons.

Thus, the investigation of the production of four more
charm-strange mesons, the $D_{s0}^*(2317)$, $D_{s1}(2460)$,
$D_{s1}(2536)$ and $D_{s2}^*(2573)$ enables an up-to-date
understanding of the production of charmed mesons, the $D$ and
$D_s$ measured in heavy ion collision experiments at the LHC. We
hope that the yields and transverse momentum distributions of the
eight charmed mesons discussed here will be helpful in revealing
the hadronization processes as well as the strangeness enhancement
for charmed hadrons, guiding the measurement of these charmed
mesons in future heavy-ion collision experiments, and thereby
contributing to a better understanding of the properties of the
quark-gluon plasma.

\section*{Acknowledgements}

This work was supported by the National Research Foundation of
Korea (NRF) grant funded by the Korea government (MSIT) No.
RS-2023-00280831 (S.C.), No. RS-2023-NR077232 (S.H.L.).

\end{document}